\documentclass[
  aps,
  prl,
  reprint,
  superscriptaddress,
  amsmath,amssymb
]{revtex4-2}

\usepackage{graphicx}
\usepackage{dcolumn}
\usepackage{bm}
\usepackage{mathtools}
\usepackage[version=4]{mhchem}
\usepackage{comment}
\usepackage{xcolor}
\usepackage[colorlinks=true,linkcolor=blue,citecolor=blue,urlcolor=blue]{hyperref}

\newcommand{\figureequivalent}{}

\begin{document}

\title{Ultralow-Field Triplon Condensation in a Spin-Ladder Magnet}

\author{Ankit Labh}
\email[Corresponding author: ]{ankit.labh@matfyz.cuni.cz}
\affiliation{Faculty of Mathematics and Physics, Department of Condensed Matter Physics, Charles University, Ke Karlovu 5, 121 16, Praha, Czech Republic}
\affiliation{Physik-Department, Technische Universität München, Garching D-85748, Germany}

\author{Ross H. Colman}
\affiliation{Faculty of Mathematics and Physics, Department of Condensed Matter Physics, Charles University, Ke Karlovu 5, 121 16, Praha, Czech Republic}

\author{Jakub \v{S}ebesta}
\affiliation{IT4Innovations Center, VSB--Technical University of Ostrava, Ostrava, Czech Republic}

\author{Noah Oefele}
\affiliation{Experimental Physics VI, Center for Electronic Correlations and Magnetism, University of Augsburg, 86159 Augsburg, Germany}

\author{Elsa Lhotel}
\affiliation{Institut Néel, CNRS, Université Grenoble Alpes, 38042 Grenoble, France}

\author{Adam Berlie}
\affiliation{ISIS, Rutherford Appleton Laboratory, Harwell Oxford, Didcot OX11 0QX, United Kingdom}

\author{Paul Steffens}
\affiliation{Institut Laue-Langevin, 71 avenue des Martyrs, CS 20156, 38042 Grenoble cedex 9, France}

\author{Oksana Zaharko}
\affiliation{PSI Center for Neutron and Muon Sciences, Forschungsstrasse 111, 5232 Villigen, PSI, Switzerland}

\author{Pascal Manuel}
\affiliation{ISIS, Rutherford Appleton Laboratory, Harwell Oxford, Didcot OX11 0QX, United Kingdom}

\author{Iurii Kibalin}
\affiliation{Institut Laue-Langevin, 71 avenue des Martyrs, CS 20156, 38042 Grenoble cedex 9, France}
\affiliation{Data Management and Scientific Computing, European Spallation Source ERIC, Asmussens Allé 305, 2800 Kongens Lyngby, Denmark}

\author{Philipp Gegenwart}
\affiliation{Experimental Physics VI, Center for Electronic Correlations and Magnetism, University of Augsburg, 86159 Augsburg, Germany}

\author{Dominik Legut}
\affiliation{Faculty of Mathematics and Physics, Department of Condensed Matter Physics, Charles University, Ke Karlovu 5, 121 16, Praha, Czech Republic}
\affiliation{IT4Innovations Center, VSB--Technical University of Ostrava, Ostrava, Czech Republic}

\author{Johanna K. Jochum}
\affiliation{Heinz Maier-Leibnitz Zentrum (MLZ), Technische Universität München, Garching D-85748, Germany}
\affiliation{Physik-Department, Technische Universität München, Garching D-85748, Germany}

\author{Alexander A. Tsirlin}
\affiliation{Felix Bloch Institute for Solid-State Physics, University of Leipzig, 04103 Leipzig, Germany}

\author{Petr \v{C}erm\'ak}
\email[Corresponding author: ]{petr.cermak@matfyz.cuni.cz}
\affiliation{Faculty of Mathematics and Physics, Department of Condensed Matter Physics, Charles University, Ke Karlovu 5, 121 16, Praha, Czech Republic}

\begin{abstract}
We realise the first ultralow-field Bose--Einstein condensation of triplons in a spin-ladder magnet, uncovering a quantum critical point at only $\mu_0H_{c1}=0.17$~T in Henmilite (\ce{Ca2Cu(OH)4[B(OH)4]2}). Unlike dimer magnets, a ladder retains extended one-dimensional correlations in its gapped parent state, making this limit strongly fluctuation dominated. 
Thermodynamic, magnetoelastic, $\mu$SR, and neutron-diffraction measurements overturn the previous assignment of zero-field antiferromagnetic order, establishing a quantum-disordered coupled-ladder parent state with persistent low-energy dynamics. The weak low-temperature anomaly instead marks a gap-controlled crossover from the correlated ladder regime into the activated quantum-disordered state.
These measurements further reveal an exceptionally asymmetric ordered dome extending to $\mu_0H_{c2}\simeq8.2$~T. 
Quantum Monte Carlo simulations for the relevant spin Hamiltonian place Henmilite just on the gapped side of the zero-field ladder-ordering instability, naturally accounting for the strong separation between the exchange and residual-gap scales and the tiny critical field.
Our findings extend ultralow-field triplon condensation beyond the dimer paradigm and establish Henmilite as a platform for controlled tuning across quantum criticality in a fluctuation-dominated spin ladder.
\end{abstract}

\maketitle


\figureequivalent

\begin{figure*}[t]
  \includegraphics[width=\textwidth]{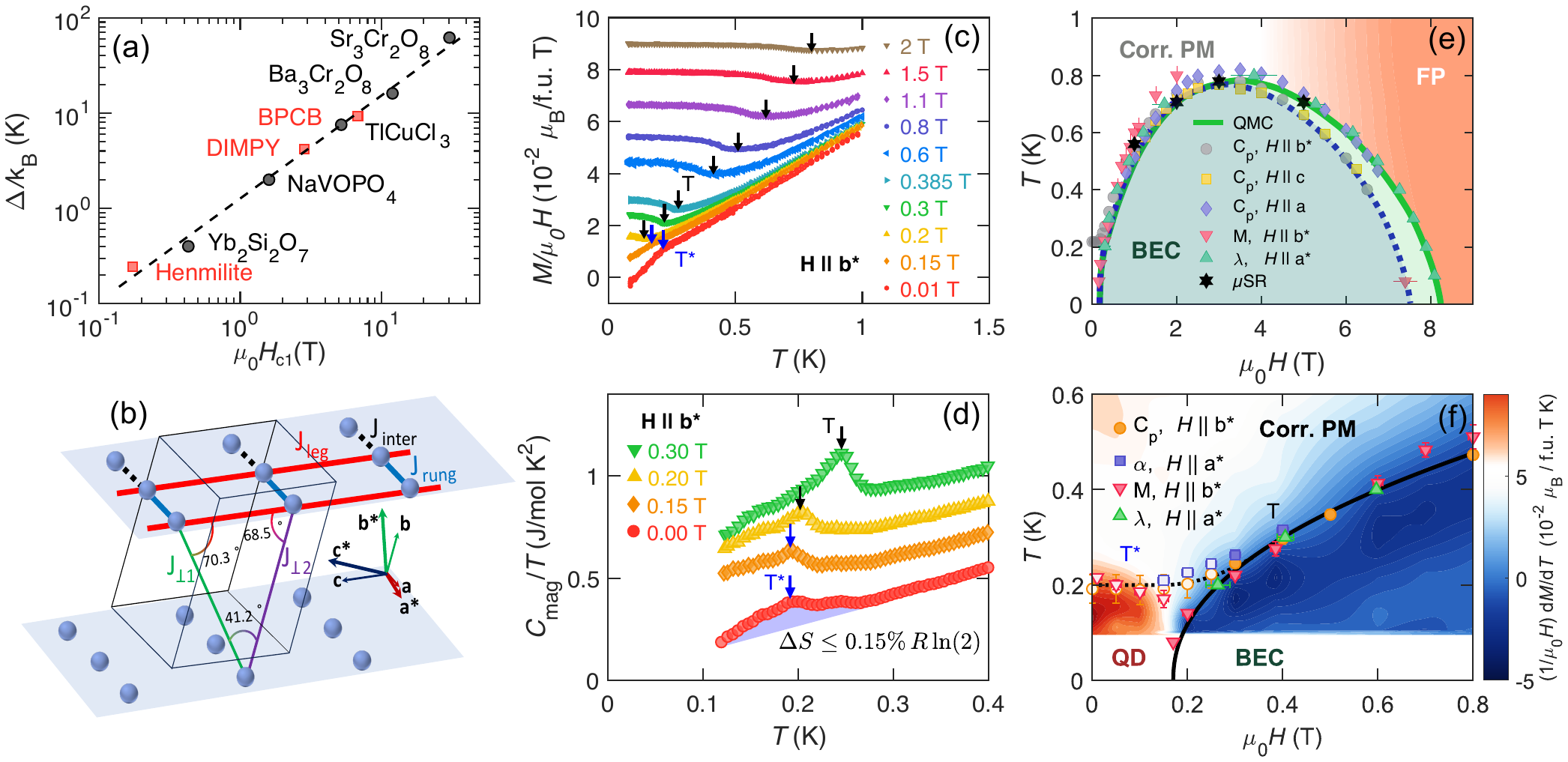}
  \caption{ 
(a) Zero-field spin gap $\Delta/k_B$ versus lower critical field
$\mu_0H_{c1}$ for representative gapped field-induced-order quantum magnets.
The dashed line marks the Zeeman gap-closing scale; red symbols denote
spin ladders and place Henmilite at the low-field end.
(b)  Crystal structure of triclinic Henmilite showing the leg
($J_{\mathrm{leg}}$), rung ($J_{\mathrm{rung}}$), and leading in-plane
interladder ($J_{\mathrm{inter}}$) exchange pathways. $J_{\perp1}$ and
$J_{\perp2}$ denote weak interlayer couplings omitted from the QMC
model.
(c) Temperature dependence of $M/\mu_0H$ for $H\parallel b^*$ between
0.01 and 2~T.
(d) Magnetic specific heat $C_{\mathrm{mag}}/T$ for
$H\parallel b^*$; the zero-field feature carries only
$\Delta S\lesssim0.15\%\,R\ln2$. Successive curves are shifted
vertically by 0.15 units for clarity.
(e) $H$--$T$ phase diagram from multiple probes. The solid green curve
is the QMC phase boundary for $H\parallel a$, while the blue dotted curve
is a guide to the eye for $H\parallel b^*$.
(f) Expanded low-field phase diagram. The solid black curve is a fit
to the field-induced phase boundary, extrapolating to
$\mu_0H_{c1}=0.17$~T at $T=0$, while the dotted line marks the
crossover scale $T^*$. The colour map shows the temperature derivative
of the magnetic susceptibility, calculated from smoothed magnetisation
data for $H\parallel b^*$.
In (e,f), labels identify the QD, BEC, correlated-paramagnetic
(corr. PM), and field polarised (FP) regimes. In (c,d,f), black arrows
and filled symbols mark field-induced phase transitions, whereas blue
arrows and open symbols denote low-field crossover scales.
  }
  \label{fig:bulk}
\end{figure*}

Quantum magnets offer unusually clean access to phase boundaries between
nonmagnetic quantum-disordered (QD) states and magnetic order. A central
example is field-induced triplon condensation in dimer magnets: on the gapped side of
such a boundary, magnetic field lowers the energy of triplet excitations
until the spin gap closes and a finite density of triplons condenses into
transverse antiferromagnetic order \citep{Nikuni_PRL_2000, Giamarchi_NatPhys_2008, Zapf_RMP_2014}. 
This field-induced ordered state is commonly described as a Bose-Einstein condensate
(BEC) of triplons, and its phase boundary forms a characteristic BEC dome 
in the field-temperature phase diagram. Canonical quantum dimer materials established 
this physics decisively, including \ce{TlCuCl3} \citep{Ruegg_Nature_2003}, 
\ce{BaCuSi2O6} \citep{Jaime_PRL_2004}, \ce{Ba3Cr2O8} \citep{Aczel_PRB_2009_Ba3Cr2O8},
and \ce{Sr3Cr2O8} \citep{Aczel_PRL_2009_Sr3Cr2O8}. Most of these benchmarks reach
the condensate only at fields of several to tens of Tesla. Lower-field examples
are rare: the weakly alternating spin-chain compound \ce{NaVOPO4} has a small gap
$\Delta_0/k_B\simeq2.4$ K and orders above $\mu_0H_{c1}\simeq1.6$ T, placing it
close to the one-dimensional quantum critical point \citep{Mukharjee_PRB_2019}.
The important ultralow-field dimer benchmark is \ce{Yb2Si2O7}, a clean quantum dimer
magnet whose entire field-induced BEC dome lies between $\mu_0H_{c1}\sim0.4$ T and
$\mu_0H_{c2}\sim1.4$ T \citep{Hester_PRL_2019_Yb2Si2O7}. 
Figure~\ref{fig:bulk}(a) summarises this progression, with
\ce{Yb2Si2O7} defining the lowest-field established benchmark among the
dimer and chain compounds.

A more fluctuation-dominated route to triplon condensation is provided by
spin-ladder compounds. Unlike nearly isolated dimers, ladders retain
extended one-dimensional correlations in their gapped parent state, while
field-induced order is governed by the balance of
$J_{\mathrm{leg}}$, $J_{\mathrm{rung}}$, and weaker interladder
couplings \citep{Giamarchi_PRB_1999,Bouillot_PRB_2011_BPCB}
(see Fig.~\ref{fig:bulk}b).
Established ladder compounds illustrate the distinct physics that emerges
from this geometry. In the strong-rung ladder
\ce{(C5H12N)2CuBr4} (BPCB), NMR, thermodynamic, and magnetoelastic
measurements established a field-tuned Tomonaga-Luttinger liquid that
develops three-dimensional order through weak interladder coupling
\citep{Klanjsek_PRL_2008_BPCB,Ruegg_PRL_2008_BPCB,
Lorenz_PRL_2008_BPCB}, while neutron spectroscopy resolved the associated
fractionalised excitation continuum
\citep{Thielemann_PRL_2009_BPCB}.
The strong-leg ladder \ce{(C7H10N)2CuBr4} (DIMPY) instead exhibits
attractive interactions and qualitatively different excitation spectra
\citep{Schmidiger_PRL_2012_DIMPY,
Schmidiger_PRL_2013_DIMPY}, while weak interladder coupling ultimately stabilises field-induced three-dimensional order \citep{Jeong_PRL_2013_DIMPY}.
These canonical gapped ladder magnets are shown in red in
Fig.~\ref{fig:bulk}a, but their gaps close only at lower critical fields of a few Tesla, well above the ultralow-field scale of \ce{Yb2Si2O7}
\citep{Watson_PRL_2001_BPCB,Hong_PRL_2010_DIMPY}.

Increasing the interladder coupling further eventually closes the gap
already at zero field and stabilises antiferromagnetic order without an
applied field. This regime is realised in \ce{Ba2CuTeO6}, where orbital
ordering generates a coupled-ladder network close to the zero-field
ordering instability
\citep{Rao_PRB_2016_BCTO,Gibbs_PRB_2017_BCTO}, and in the planar ladder
\ce{C9H18N2CuBr4} (DLCB), where proximity to the same instability gives
rise to a long-lived Higgs amplitude mode and field-induced quasiparticle
decay
\citep{Hong_PRB_2014_DLCB,Hong_NatPhys_2017_DLCB,
Ying_PRL_2019_DLCB,Hong_NatCommun_2017_DLCB}.
Because these compounds are already ordered at zero field, they have no
finite gap-closing field and therefore do not appear in the
gap--$H_{c1}$ comparison of Fig.~\ref{fig:bulk}a.

Taken together, these examples reveal a largely unexplored regime:
the ultralow-field accessibility demonstrated by \ce{Yb2Si2O7} has not
yet been realized in a spin ladder magnet. The canonical gapped ladders BPCB
and DIMPY retain gaps that close only at fields of a few Tesla
\citep{Watson_PRL_2001_BPCB,Hong_PRL_2010_DIMPY}, whereas
\ce{Ba2CuTeO6} and DLCB have already crossed into antiferromagnetic
order at zero field. 
A ladder that remains gapped but lies just short of this instability
would combine an ultralow critical field with the extended
one-dimensional correlations and dimensional crossover intrinsic to
ladder magnets. Its tiny residual gap would additionally
make the zero-field ground state particularly sensitive to nonmagnetic
tuning parameters, offering a route across the ordering instability
through pressure or chemical substitution.

Henmilite, \ce{Ca2Cu(OH)4[B(OH)4]2}, is a naturally occurring $S=1/2$ Cu 
mineral whose exchange network has been proposed to form coupled spin ladders (Fig.~\ref{fig:bulk}b),
but whose low-temperature state was previously assigned to antiferromagnetic
order \citep{yamamoto2021quantum}. We show instead that its bulk zero-field
state is a quantum-disordered coupled-ladder, lying on the gapped
side of a nearby ordering instability. The gap is exceptionally small: 
the fitted lower critical field is only $\mu_0H_{c1}=0.17$ T, corresponding 
to a Zeeman scale below that of the ultralow-field dimer benchmark 
\ce{Yb2Si2O7} \citep{Hester_PRL_2019_Yb2Si2O7} and more than an order
of magnitude below the established spin-ladder compound DIMPY \citep{Schmidiger_PRL_2012_DIMPY}.
A field-induced ordered dome then emerges from this quantum-disordered
parent state, consistent with triplon condensation in a coupled ladder.
Below $\mu_0H_{c1}$, the weak low-temperature anomaly is instead attributed to
a gap-controlled crossover from the near-critical correlated regime into
the activated quantum-disordered regime.

Low-temperature DC magnetisation, thermodynamic, and magnetoelastic
measurements identify the low-field crossover and an ultralow-field 
ordered dome emerging at $\mu_0H_{c1}=0.17$~T.
Together with $\mu$SR, these measurements rule out conventional order in zero field
and reveal persistent low-energy dynamics.
Quantum Monte Carlo (QMC) calculations for a microscopic coupled-ladder
model containing the three leading exchange couplings account for the tiny
critical field by placing Henmilite on the gapped side of a nearby
ladder-ordering instability.
Neutron diffraction then supports the field-induced magnetisation scale
microscopically: the field-dependent intensity at the nuclear Bragg
reflection $(4~0~1)$ follows the QMC magnetisation scaling.

All measurements were performed on selected natural Henmilite crystals
obtained from a commercial mineral specimen. Because the specimen also
contained non-magnetic Olshanskyite and Calcite, individual blue crystallites were
optically selected and screened by laboratory X-ray diffraction before use.
Laboratory Xray powder diffraction refined the triclinic Henmilite structure with lattice parameters obtained at room temperature $a$~=~11.537(1)\AA\,, $b$~=~7.987(1)\AA, $c$~=~5.649(1)\AA, $\alpha$~=~109.625(2)$^\circ$, $\beta$~=~91.519(2)$^\circ$, $\gamma$~=~83.678(1)$^\circ$, consistent with the published structure \citep{yamamoto2021quantum}. Repeated measurements on independently selected single crystals, together with powder diffraction on material prepared from many selected crystallites, gave the same Henmilite phase.
Controlled laboratory growth was attempted but did not yield suitable crystals; details are given in the Supplemental Material \citep{supplement}.

We next turn to the low-temperature magnetic response, where the
magnetisation most clearly reveals qualitatively distinct behaviour
below and above $H_{c1}$. In $M/\mu_0H$ for 
$H\parallel b^*$ (Fig.~\ref{fig:bulk}c),
a clear transition can be followed down to fields of order $0.2$~T.
Below this field scale, the sharp minimum gives way to a much weaker kink,
seen as a steeper downturn on cooling. This distinction is first apparent
at $\mu_0H=0.15$~T, where no magnetic transition is resolved down to
$80$~mK, whereas higher fields show clear anomalies associated with the
ordered dome. 
Magnetic specific heat $C_{\mathrm{mag}}/T$
(Fig.~\ref{fig:bulk}d) shows the same evolution: the clear anomalies
at $0.2$ and $0.3$~T, associated with the field-induced ordered phase,
give way to much weaker and broader features at $0$ and $0.15$~T.
The weak zero-field feature, which was previously associated with the onset of magnetic order~\cite {yamamoto2021quantum}, is therefore
reproduced in our data, but the associated excess entropy is only
$\Delta S\lesssim0.15\%\,R\ln2$ per Cu (shaded area in
Fig.~\ref{fig:bulk}d), showing that only a minute residual part of the
spin entropy is involved. 
Its negligible shift in temperature between zero field and $\mu_0H=0.15$~T, together with the absence of a corresponding transition in the magnetisation, shows that this weak, broad feature does not mark the onset of conventional bulk magnetic order. The characteristic temperatures of these weak low-field features are denoted by $T^*$ in Figs.~\ref{fig:bulk}c,d.
Thermal-expansion and magnetostriction measurements show a similar low-field evolution; representative traces and details of the entropy extraction are given in the Supplemental Material~\citep{supplement}. The $T^*$ values extracted from magnetisation, specific heat, and magnetoelastic measurements are shown as open symbols in Fig.~\ref{fig:bulk}f.

The resulting bulk $H$--$T$ phase diagram is summarised in
Fig.~\ref{fig:bulk}e, with the low-field region expanded in
Fig.~\ref{fig:bulk}f. Measurements for several field orientations show
moderate anisotropy but the same overall topology and pronounced
asymmetry of the field-induced dome. The BEC phase extends to
$\mu_0H_{c2}\simeq8.2 \pm 0.04$~T, as determined from magnetostriction
(see Supplemental Material \citep{supplement}), above which the system
enters the FP regime.
On the low-field side of the dome, by contrast, the open symbols represent the $T^*$ crossover scales below $H_{c1}$ and are therefore excluded from the phase-boundary fit. Fitting only the field-induced transition temperatures yields
$\mu_0H_{c1}=0.17 \pm 0.02$~T, giving an exceptionally small width-normalised
onset field, $H_{c1}/(H_{c2}-H_{c1})\simeq0.02$.    
Using the average ESR-derived $g$ factor, $g=2.15$
\citep{Hayashi_JPSCP_2023}, the lower
critical field corresponds to a Zeeman gap $\Delta/k_B\sim0.245$~K, comparable to the $T^*$ scale.

The low-field anomaly at $T^*$ reflects the strong separation between the exchange
and residual-gap scales. Short-range ladder correlations develop on the
few-Kelvin exchange scale \citep{yamamoto2021quantum}, producing the
`corr. PM' regime, whereas the residual gap is only
$\Delta/k_B\simeq0.245$~K. Cooling through $T^*$ suppresses the population of the lowest triplet state and marks a crossover into the activated QD regime.
This crossover is seen most clearly in the magnetisation
(Fig.~\ref{fig:bulk}c), where the decrease of $M/\mu_0H$ accelerates
below $T^*$, while weak thermodynamic and dilatometric anomalies occur
at comparable temperatures.

\begin{figure}[tb]
  \includegraphics[width=\columnwidth]{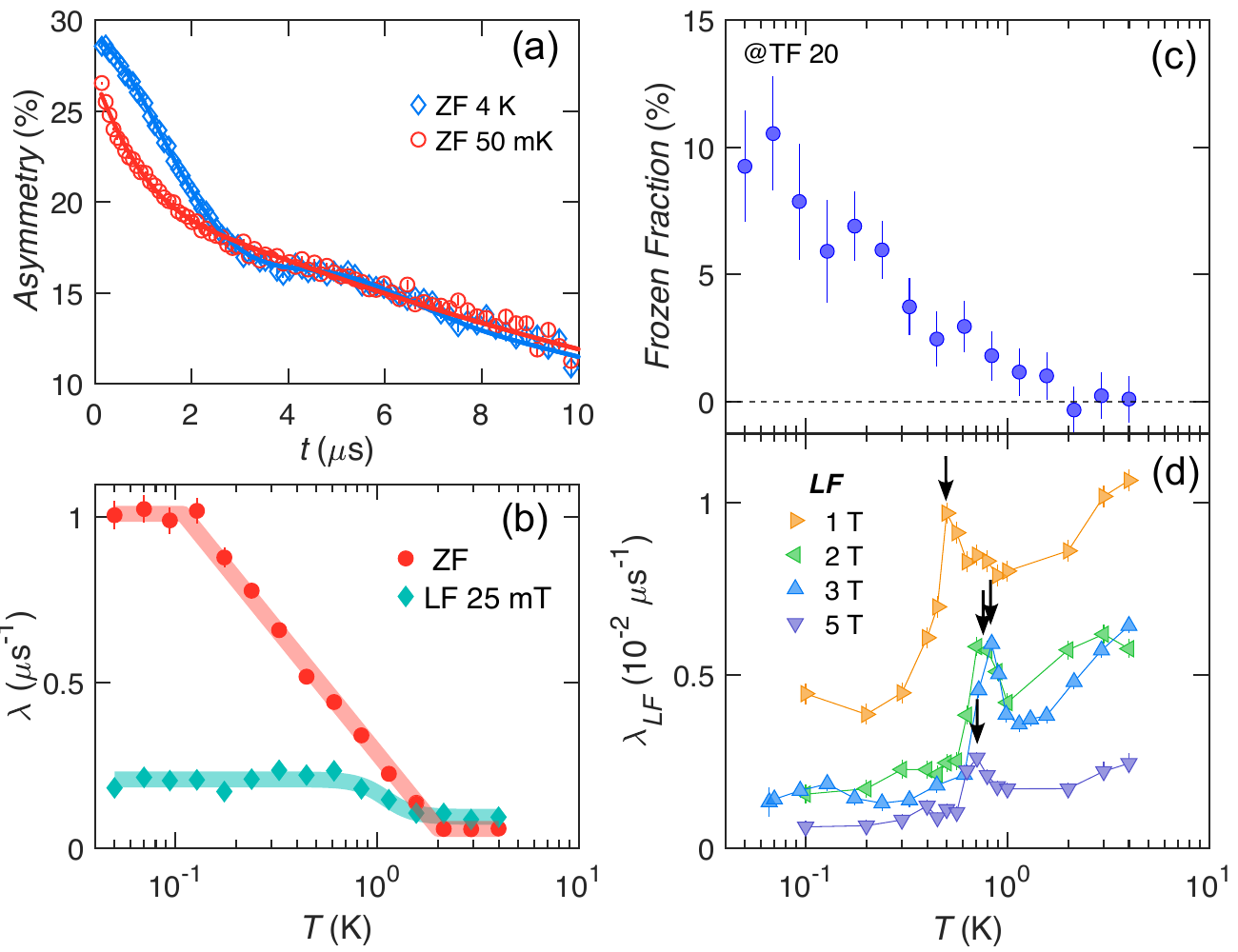}
  \caption{
(a) Zero-field $\mu$SR spectra at 4 K and 50 mK. The 50 mK spectrum acquires an additional electronic relaxation channel, demonstrating slow magnetic fluctuations, but shows neither spontaneous oscillations nor the loss of asymmetry characteristic of conventional static bulk antiferromagnetism.
(b) Temperature dependence of the electronic relaxation rate $\lambda$
in zero field (ZF) and at longitudinal field (LF) $\mu_0H_{\rm LF}=25$~mT, chosen to decouple
the quasi-static nuclear contribution from electronic relaxation.
(c) Frozen fraction on the $\mu$SR time scale. Only a small fraction becomes frozen at low temperature, unlike the
magnetically ordered low-temperature response reported for the
coupled-ladder compound \ce{Ba2CuTeO6} \citep{Glamazda_PRB_2017_BCTO}.
(d) Temperature dependence of the electronic relaxation rate $\lambda$
in longitudinal fields $\mu_0H_{\rm LF}=1$--5~T. Sharp anomalies,
indicated by arrows, mark the field-induced phase boundary and are
included in Fig.~\ref{fig:bulk}e.
  }
  \label{fig:musr}
\end{figure}

To test whether the low-field anomaly could nevertheless reflect static
magnetic order, we performed $\mu$SR on the
HiFi instrument at the ISIS Neutron and Muon Source
\citep{HiFi, ISIS_RB2310640_HIFI}. If the low-field feature marked conventional
full-volume antiferromagnetic order, the onset of static internal fields would
be expected to produce spontaneous oscillations or a substantial loss
of initial asymmetry due to unresolved rapid depolarisation.

Instead, the zero-field muon asymmetry relaxes monotonically down to
$50$~mK [Fig.~\ref{fig:musr}(a)], without such signatures.
The low-temperature spectrum is nevertheless clearly distinct from the
4~K response: the additional early-time depolarisation for
$t\lesssim2~\mu\text{s}$ is captured by an electronic relaxation channel,
consistent with spin fluctuations slowing into the $\mu$SR time window
($10^6$--$10^9~\text{s}^{-1}$).
The electronic relaxation rate $\lambda$ increases on cooling and 
saturates at low temperature (Fig.~\ref{fig:musr}b).
A weak longitudinal field suppresses much of the
low-temperature relaxation but leaves a residual slow channel, demonstrating
that low-energy electronic spin dynamics persist down to base temperature (see Supplemental Material \citep{supplement}).

The fitted frozen fraction, calculated from weak transverse-field (TF) measurements,
remains below approximately $10\%$
(Fig.~\ref{fig:musr}c) and therefore far from representing the bulk.
This fraction should not be interpreted as a direct measure of impurity
concentration, since dilute defects may perturb a larger surrounding spin volume.
In an applied field, $\lambda(T)$ develops sharp maxima at temperatures
that agree with the phase boundary independently determined from the bulk
probes (Fig.~\ref{fig:musr}d and Fig.~\ref{fig:bulk}e). These maxima are consistent with critical
slowing down as the electronic fluctuation time scale enters the $\mu$SR
window on approaching the field-induced ordered phase.
Thus $\mu$SR rules out conventional full-volume static antiferromagnetic
order at zero field, reveals persistent low-energy spin dynamics, and
independently tracks the onset of the field-induced ordered phase.

\begin{figure}[tb]
  \includegraphics[
    width=\columnwidth,
    height=0.72\textheight,
    keepaspectratio
  ]{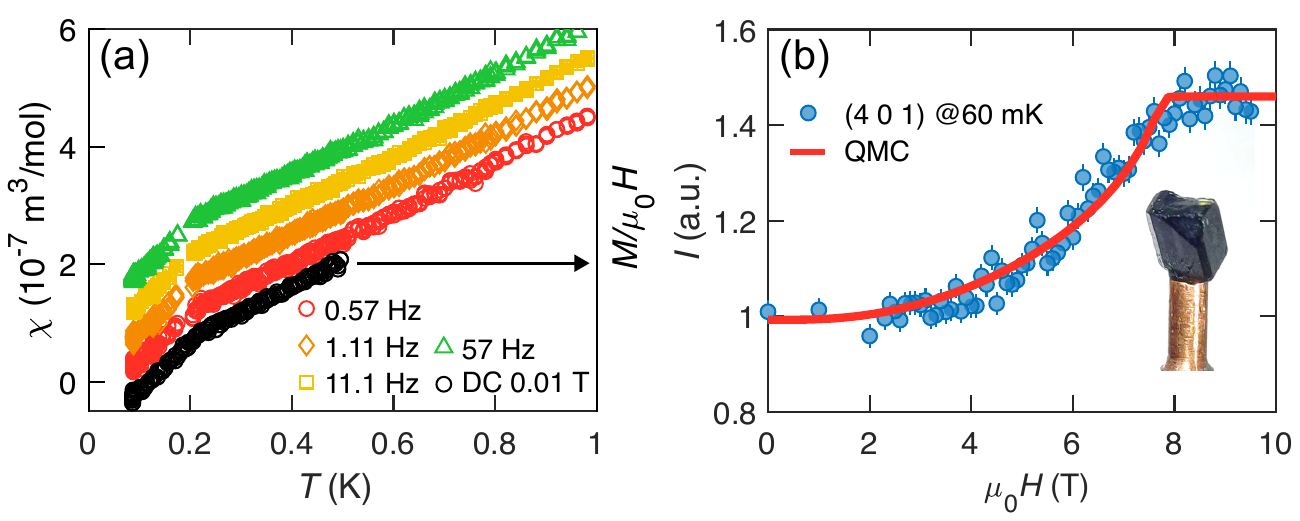}
  \caption{
(a) AC susceptibility measured at several frequencies in zero DC field, together with the DC susceptibility at $\mu_0H=0.01$~T. The AC curves are shifted vertically for clarity; the absence of a measurable frequency shift of the low-temperature feature argues against a conventional spin-glass-like freezing transition.
(b) Field dependence of the intensity at the nuclear Bragg reflection $(4,0,1)$ measured at 60~mK. The field-induced intensity follows the squared uniform magnetisation from the coupled-ladder QMC calculation (red solid line), providing a microscopic check of the field-induced magnetisation scale. The inset shows the 76~mg natural Henmilite crystal used for the neutron-diffraction measurements.
  }
  \label{fig:constraints}
\end{figure}

AC susceptibility measured across the weak low-field feature shows no
measurable frequency shift within the experimental window
(Fig.~\ref{fig:constraints}a). No dissipative response $\chi''$ is
detected up to 1.2~K, providing no evidence for conventional spin-glass
freezing over the measured frequency range
\citep{Malinowski_PRB_2011_SG}.
Neutron diffraction provides a complementary microscopic probe of the
field-induced response (Fig.~\ref{fig:constraints}b). Measurements
were performed on D10 at ILL \citep{ILL_DATA_5_41_1274}, ZEBRA at PSI \citep{PSI}, WISH at the ISIS Neutron and Muon source \citep{WISH, ISIS_RB2420429_WISH_1, ISIS_RB2400089_WISH_2}, and ThALES at ILL \citep{ILL_DATA_TEST_3445} using the largest
available Henmilite crystal ($76$~mg). Neutron absorption by natural
boron and the large incoherent background from hydrogen strongly limit
the magnetic sensitivity, with ThALES providing the best
signal-to-background ratio using an optimised elastic configuration with
$k_i=k_f=1.45$~\AA$^{-1}$ and $60'$ post-sample collimation
\citep{supplement}.
At 60~mK, the integrated intensity of the
nuclear $(4~0~1)$ reflection increases with field and follows the
squared uniform magnetisation extracted from the QMC calculation
(Fig.~\ref{fig:constraints}b), demonstrating a field-induced magnetic
contribution at $\mathbf{k}=0$.
Extended counting at the candidate antiferromagnetic reflection
$(1\,0\,\tfrac12)$, selected from a model-motivated
$\mathbf{k}=(0,0,\tfrac12)$ structure, revealed no statistically
significant field-induced intensity. The propagation vector and
staggered moment therefore remain unresolved.

\begin{figure}[tb]
  \includegraphics[width=\columnwidth]{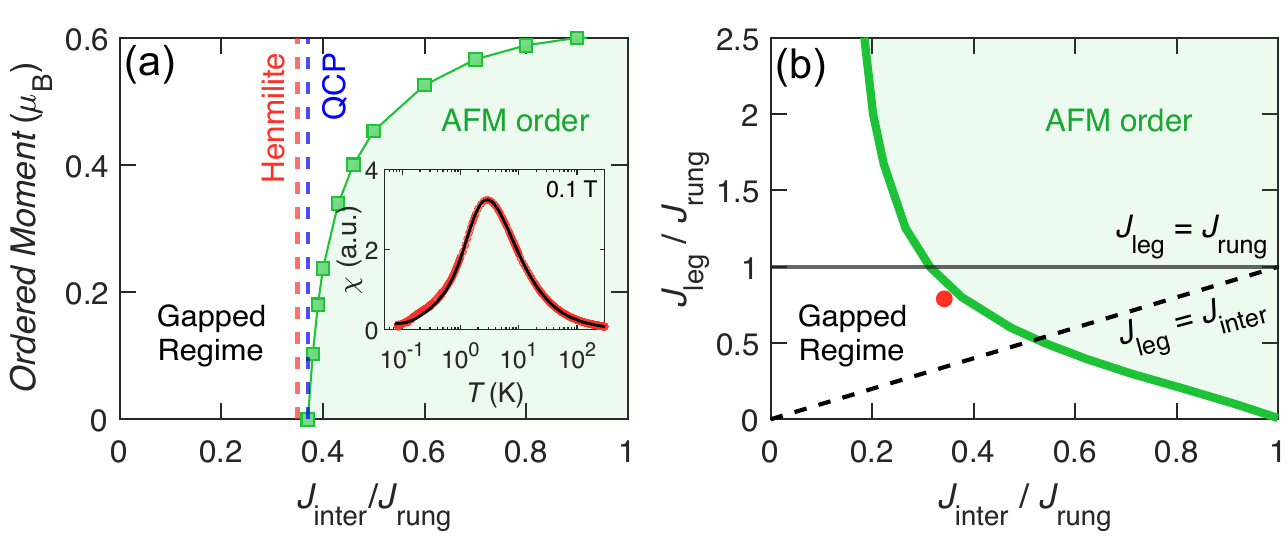}
  \caption{
(a) QMC analysis of the microscopic coupled-ladder model for fixed $J_{\mathrm{leg}}/J_{\mathrm{rung}}=0.8$. The ordered moment turns on rapidly above the zero-field critical interladder coupling of $J_{\rm inter}/J_{\rm rung}$ = 0.37. The fitted exchange parameters of Henmilite lie just below this threshold, on the gapped side of the instability. The inset shows the corresponding dc susceptibility measured at 0.1 T together with QMC fit for $J_{\mathrm{rung}}=3.8$~K, $J_{\mathrm{leg}}/J_{\mathrm{rung}}=0.8$, and $J_{\mathrm{inter}}/J_{\mathrm{rung}}=0.35$.
(b) Generic zero-field phase diagram for coupled spin ladders, adapted from Ref.~\cite{Matsumoto_PRB_2001}. The diagram separates the spin-gapped ladder regime from antiferromagnetic long-range order as a function of leg, rung, and interladder exchange ratios. The position inferred for Henmilite lies close to the ordering boundary, consistent with a quantum-disordered parent state whose residual gap is closed by a field of only $0.17$ T.
  }
  \label{fig:qmc}
\end{figure}

To connect the ultralow critical field and field-induced response to
microscopic exchange parameters, we model Henmilite using the $S=1/2$
coupled-ladder model retaining the three leading exchange couplings,
$J_{\mathrm{leg}}$, $J_{\mathrm{rung}}$, and $J_{\mathrm{inter}}$. Quantum Monte Carlo calculations for this model were first constrained by the
low-field dc susceptibility, which fixes the dominant exchange scale and
the leg--rung ratio. A single parameter set, $J_{\mathrm{leg}}=3.0$~K, $J_{\mathrm{rung}}=3.8$~K, and $J_{\mathrm{inter}}=1.3$~K, reproduces the
broad susceptibility maximum (Fig.~\ref{fig:qmc}a, inset) along with the magnetisation and field-induced
neutron intensity discussed above. The resulting ratios,
$J_{\mathrm{leg}}/J_{\mathrm{rung}}\simeq0.8$ and
$J_{\mathrm{inter}}/J_{\mathrm{rung}}\simeq0.35$, place Henmilite just below
the zero-field ordering threshold. In the QMC scan of the ordered
moment, the onset of magnetic order occurs near
$J_{\mathrm{inter}}/J_{\mathrm{rung}}\simeq0.37$, so the fitted Henmilite point
lies on the gapped side but very close to the instability
(Fig.~\ref{fig:qmc}a). This proximity naturally accounts for the strong separation between the
dominant $J\sim3$--$4$~K scale and the much smaller residual gap
$\Delta/k_B\simeq0.245$~K underlying the low-field crossover $T^*$.

First-principles calculations independently support this microscopic
description. Both our DFT calculations and the earlier DFT
analysis~\citep{yamamoto2021quantum} of Henmilite yield antiferromagnetic exchanges of the order a few
Kelvin, and the summed exchange scale is similar to that of the
QMC-constrained model. DFT further shows that the leading interladder coupling
$J_{\mathrm{inter}}$ connects neighbouring ladders in-plane, whereas
$J_{\perp1}$ and $J_{\perp2}$ are much weaker interlayer couplings
and are therefore neglected in the QMC calculations. Modest differences
in the distribution of exchange strength among the three leading
pathways are common in low-dimensional quantum magnets, where individual
exchange values are sensitive to structural details and computational
methodology
\citep{Jeschke2011Azurite,Tsirlin2010BiCu2PO6}.
Importantly, all approaches agree on the microscopic coupled-ladder
topology and the relevant energy scale of few-Kelvin. The DFT results, the previous DFT
values~\citep{yamamoto2021quantum}, and the QMC parameter set are compared in the Supplemental
Material \citep{supplement}.

This placement is summarised in Fig.~\ref{fig:qmc}b, where the
fitted exchange ratios are compared with the generic zero-field phase
diagram of coupled spin ladders \citep{Matsumoto_PRB_2001}. The diagram
separates a spin-gapped, quantum-disordered ladder regime from
antiferromagnetic long-range order as a function of the leg, rung, and
interladder exchange ratios. 
Henmilite falls on the gapped side of this boundary, whereas ordered
coupled-ladder compounds such as \ce{LaCuO_{2.5}}
\citep{Normand_PRB_1997} and \ce{Ba2CuTeO6}
\citep{Macdougal_PRB_2018_BCTO} lie close to the same instability
from the antiferromagnetic side. 
The proximity to this boundary accounts for the very small lower critical field, $\mu_0H_{c1}=0.17$~T, at which the zero-field parent state enters the triplon-BEC dome.
The extreme dome asymmetry provides a complementary phenomenological
measure of this near-criticality. Applying the effective-mass analysis
developed for the canonical spin-1 BEC magnet
\ce{NiCl2.4SC(NH2)2} (DTN) \citep{Kohama_PRL_2011_DTN} yields
$m^*/m\simeq24$, where $m^*$ and $m$ denote the effective masses near
the lower and upper critical fields, respectively, compared with
$m^*/m\simeq3.2$ in DTN. Although this mapping is phenomenological,
the unusually large ratio is consistent with strong low-field
quasiparticle renormalisation; details and assumptions are given in the
Supplemental Material \citep{supplement}.

The resulting picture is that Henmilite realises the missing gapped-side
limit of a coupled spin ladder sitting exceptionally close to the magnetically
ordered phase. The previously assigned zero-field antiferromagnetic state
\citep{yamamoto2021quantum} should be revised in favour of a
quantum-disordered parent state with persistent low-energy dynamics.
Short-range ladder correlations develop on the few-Kelvin exchange
scale, while cooling through $T^*\sim\Delta/k_B$ suppresses the remaining
triplet population and produces the crossover into the activated QD
regime. 
The present data exclude conventional full-volume antiferromagnetic LRO
as the origin of the zero-field anomaly. In the absence of resolved
symmetry breaking or an unambiguous thermodynamic singularity, we
therefore identify $T^*$ with the gap-controlled crossover. A true ordered
phase appears only after the residual gap closes at
$\mu_0H_{c1}=0.17$~T. This onset lies far below those of the canonical
spin ladders BPCB and DIMPY and even below the ultralow-field dimer
benchmark \ce{Yb2Si2O7} (Fig.~\ref{fig:bulk}a). Nevertheless, the
field-induced phase expands into a broad BEC dome extending over several
Tesla and reaching temperatures of order $1$~K.

Henmilite therefore brings spin-ladder quantum criticality into a regime
where both sides of the boundary are accessible to thermodynamic, local,
and diffraction probes. Its position just inside the gapped regime,
together with the few-Kelvin exchange scale and low-symmetry ladder
network, makes it a natural target for pressure tuning and chemical substitution. Pressure drives the ordered coupled ladder
DLCB across a quantum phase transition into a quantum-disordered state
\citep{Hong_NatCommun_2022_DLCB}, while chemical substitution strongly
modifies the coupled-ladder ground state of \ce{Ba2CuTeO6}
\citep{Badola_PRB_2024_BCTO}. Starting from the opposite, gapped side
with an exceptionally small residual gap, Henmilite offers the rare
prospect of crossing the zero-field ladder-ordering instability by
nonmagnetic tuning.


\begin{acknowledgments}
We thank Nikolaos Biniskos (ILL) for useful discussions.  This work was supported by the Czech Science Foundation GAČR under the Junior Star Grant No. 21-24965M (MaMBA) and the bilateral Czech-Bavarian projects AQuaMaRINe (BTHA-JC-2022-34) and BaCQuERel (project No. LUABA24056). 
DL and JS acknowledge the computational resources by the project e-INFRA CZ (ID:90254) and DL also project QM4ST No. CZ.02.01.01/00/22\_008/0004572 by the Ministry of Education, Youth and Sports of the Czech Republic.
Single crystal characterisation and bulk properties measurement, except for low temperature magnetisation and thermal expansion, were performed in MGML (mgml.eu), which is supported within the program of Czech Research Infrastructures (project no. LM2023065). Work at the University of Augsburg was supported by the Bavarian-Czech Academic Agency (Project no. BTHA-JC-2024-15) and by the German Research Foundation (DFG) through TRR360 (Project no. 492547816). This work is based on experiments performed at the Swiss spallation neutron source SINQ, Paul Scherrer Institute, Villigen, Switzerland. Experiments at the ISIS Neutron and Muon Source were supported by beamtime allocation RB2310640, RB2420429 and RB2400089 from the Science and Technology Facilities Council; data are available here: \citep{ISIS_RB2310640_HIFI,ISIS_RB2420429_WISH_1,ISIS_RB2400089_WISH_2}. We thank Institut Laue-Langevin for the beamtime at D10 and ThALES instruments.
\end{acknowledgments}

\noindent\textit{Data availability.—}
Raw data and figure-generation MATLAB code will be publicly available on Figshare upon publication \cite{Labh_Figshare_2026}.

\nocite{Paulsen_2001,Kuchler_RSI_2017,Kunisawa_ActaCrysE_2026,fullprof,Scheie_JLTP_2018,Mantid,wimda}
\nocite{Lancaster_PRL_2007,LORD2000495,Fak_PRL_2012,Tustain_NPJQM_2020,Berlie_PRB_2022}
\nocite{Kresse_VASP_r96,Kresse_VASP_r99,PBE_r96,Bloechl_corr_tetrahedron,DFT_D3_IVDW11,MaxlocWF,Wannier90,TB2J,LKAG,Todo_PRL_2001,Alet_PRE_2005,Albuquerque_ICM_2007,Harada_PRB_1997}
\nocite{Samulon_PRL_2009,Flynn_PRL_2021_Yb2Si2O7,Feng_PRB_2023_Yb2Si2O7,Zhang_PRB_2013}

\bibliography{bibliography}

\end{document}


\title{\textbf{Supplemental Material for: Ultralow-Field Triplon Condensation in a Spin-Ladder Magnet}}

\author{Ankit Labh$^{1,2}$, Ross H. Colman$^{1}$, Jakub \v{S}ebesta$^{3}$, Noah Oefele$^{4}$, Elsa Lhotel$^{5}$, Adam Berlie$^{6}$, Paul Steffens$^{7}$, Oksana Zaharko$^{8}$, Pascal Manuel$^{6}$, Iurii Kibalin$^{7,9}$, Philipp Gegenwart$^{4}$,  Dominik Legut$^{1,3}$,  Johanna K. Jochum$^{10,2}$, Alexander A. Tsirlin$^{11}$, Petr \v{C}erm\'ak$^{1}$}
\thanks {corresponding author: \\petr.cermak@matfyz.cuni.cz; ankit.labh@matfyz.cuni.cz}

\affiliation{
 $^{1}${Faculty of Mathematics and Physics, Department of Condensed Matter Physics, Charles University, Ke Karlovu 5, 121 16, Praha, Czech Republic}
 $^{2}${Physik-Department, Technische Universität München, Garching D-85748, Germany} 
 $^{3}${IT4Innovations Center, VSB--Technical University of Ostrava, Ostrava, Czech Republic}
 $^{4}${Experimental Physics VI, Center for Electronic Correlations and Magnetism, University of Augsburg, 86159 Augsburg, Germany}
 $^{5}${Institut Néel, CNRS, Université Grenoble Alpes, 38042 Grenoble, France}
 $^{6}${ISIS, Rutherford Appleton Laboratory, Harwell Oxford, Didcot OX11 0QX, United Kingdom}
 $^{7}${Institut Laue-Langevin, 71 avenue des Martyrs, CS 20156, 38042 Grenoble cedex 9, France}
 $^{8}${PSI Center for Neutron and Muon Sciences, Forschungsstrasse 111, 5232 Villigen, PSI, Switzerland}
 $^{9}${Data Management and Scientific Computing, European Spallation Source ERIC, Asmussens Allé 305, 2800 Kongens Lyngby, Denmark}
 $^{10}${Heinz Maier-Leibnitz Zentrum (MLZ), Technische Universität München, Garching D-85748, Germany}
$^{11}${Felix Bloch Institute for Solid-State Physics, University of Leipzig, 04103 Leipzig, Germany}
}

\maketitle

\onecolumngrid 

\section{EXPERIMENTAL SECTION}

\textbf{Magnetisation}: Low-temperature direct current (DC) and alternating current (AC) magnetisation measurements were conducted in two purpose-built SQUID magnetometers equipped with a dilution refrigerator reaching a base temperature of $T \approx 70\text{--}80\text{ mK}$ \cite{Paulsen_2001} across two distinct experimental runs. To ensure optimal thermal contact and stability, the single crystal Henmilite sample of mass 10 mg was mounted using Apiezon N grease between two parallel copper plates along the magnetic field direction. Two distinct magnetometer probes were employed within the same dilution unit: a high-sensitivity, low-field instrument limited to an applied magnetic field of $\mu_0 H \le 0.385\text{ T}$, and a high-field instrument operating up to $8\text{ T}$. The DC magnetisation data were acquired via an extraction technique without explicit background subtraction. The AC susceptibility was recorded in relative mode and was rescaled to the absolute susceptibility obtained using the extraction method. At zero DC bias field, the excitation frequency was varied between $0.21\text{ Hz}$ and $211\text{ Hz}$, whereas under an applied DC field, the maximum frequency was restricted to $2.11\text{ Hz}$. The peak amplitude of the AC drive field was maintained at $H_{\text{AC}} = 3.32\text{ Oe}$. Complementary magnetisation measurements were performed using the Vibrating Sample Magnetometer (VSM) option in commercial Quantum Design $^4\text{He}$ PPMS and $^3\text{He}$-enabled MPMS3 systems.

\textbf{Magnetoelastic}: Ultrahigh-resolution thermal expansion and magnetostriction measurements were carried out in a dilution refrigerator using a miniaturised capacitive dilatometer~\cite{Kuchler_RSI_2017} on a single-crystal Henmilite sample mounted in a dedicated dilatometry cell, with the magnetic field $H\parallel a^*$. The sample was mounted between two parallel copper plates; hence, a suitable crystal with two parallel facets was chosen. For other measurements, the experimental details are provided in their respective sections.


\section{CRYSTAL GROWTH}
The growth of synthetic Henmilite, \ce{Ca2Cu(OH)4[B(OH)4]2}, single crystals would be a key prerequisite for a detailed investigation of its low-temperature dynamics by inelastic neutron scattering. This is particularly important because natural Henmilite contains both boron and hydrogen, which are intrinsically unfavourable for neutron experiments because of neutron absorption from naturally abundant boron and the large incoherent background from hydrogen. Larger, compositionally controlled synthetic crystals would therefore substantially improve the feasibility of neutron spectroscopy. Because no targeted synthetic single-crystal growth protocol for this mineral was available, our exploratory synthesis routes were modelled on conventional hydrothermal methodologies often used for natural mineral analogues and frustrated quantum magnets, including Herbertsmithite and Kapellasite.

Initial synthesis attempts focused on systematic variations of the stoichiometric precursor ratios and absolute solute concentrations using calcium hydroxide \ce{Ca(OH)2}, copper(II) oxide \ce{CuO}, and boric acid \ce{B(OH)3} with deionised water as the aqueous medium. In the primary exploratory matrix, the molar concentration of $\text{B(OH)}_3$ was systematically varied while maintaining a fixed \ce{Ca(OH)2}: \ce{CuO} ratio. In a subsequent series, the concentration of the $\text{CuO}$ precursor was varied against a constant excess of \ce{B(OH)3}. For each experimental run, the aqueous precursor mixtures were encapsulated within polytetrafluoroethylene (PTFE) ampoules and placed inside protective brass outer jackets within hydrothermal digestion vessels (Parr 4749 General Purpose Vessel) to safely contain the autogenous pressure generated at elevated temperatures. The autoclaves were heated in a conventional oven to a maximum temperature of $220\,^\circ\text{C}$ to drive dissolution, followed by a slow-cooling profile to promote crystallisation. The resulting solid products were washed repeatedly with deionised water to dissolve and remove excess unreacted boric acid, followed by paper filtration. Despite extensive variation in precursor ratios and solvent volumes, no Henmilite phase was detected, including polycrystalline material, in the recovered products.

As an alternative approach, a two-zone thermal gradient transport method was employed. The precursor materials were loaded into a thick-walled quartz tube, and the aqueous solvent was introduced carefully to ensure that the precursor materials initially remained localised at one end of the tube. The ampoule was subsequently flame-sealed under vacuum. The sealed quartz tube was placed in a two-zone horizontal furnace configured with a static temperature gradient: the source zone containing the precursors was maintained at $220\,^\circ\text{C}$, while the growth zone was kept at $170\,^\circ\text{C}$. This temperature gradient was continuously maintained for 21~days to provide a sustained thermodynamic driving force for dissolution, mass transport, and subsequent crystallisation at the cooler end. Upon completion of the cycle, visual inspection and X-ray diffraction analysis confirmed that Henmilite crystals were not obtained by this method either. After these unsuccessful attempts, all subsequent experiments were based on naturally grown specimens obtained from Fuka mines in Japan. However, a recent study reports that a powder sample of Henmilite can be obtained by a similar ammonia-evaporation method using Ca(OH)$_2$~\cite{Kunisawa_ActaCrysE_2026}, suggesting a possible starting point for future targeted growth, although an optimised Henmilite single-crystal growth protocol has not yet been established.

\section{X-ray}

X-ray powder diffraction was performed on a powder sample prepared by crushing single crystals of Henmilite and measured at room temperature on an Empyrean advanced X-ray diffractometer with Cu-K$\alpha$ radiation. Henmilite's lattice parameters were refined by the Rietveld method using FullProf \citep{fullprof}, yielding $a = 11.53711(18)$~\AA, $b = 7.98727(13)$~\AA, $c = 5.64921(10)$~\AA, $\alpha = 109.6253(11)^\circ$, $\beta = 91.5194(11)^\circ$, and $\gamma = 83.6784(10)^\circ$. These values are in good agreement with the previous report by Yamamoto \textit{et al.}~\citep{yamamoto2021quantum}.

\begin{figure}[htbp]
    \centering
    \graphicspath{{figures_supplement/}}
    \includegraphics[width=1\linewidth]{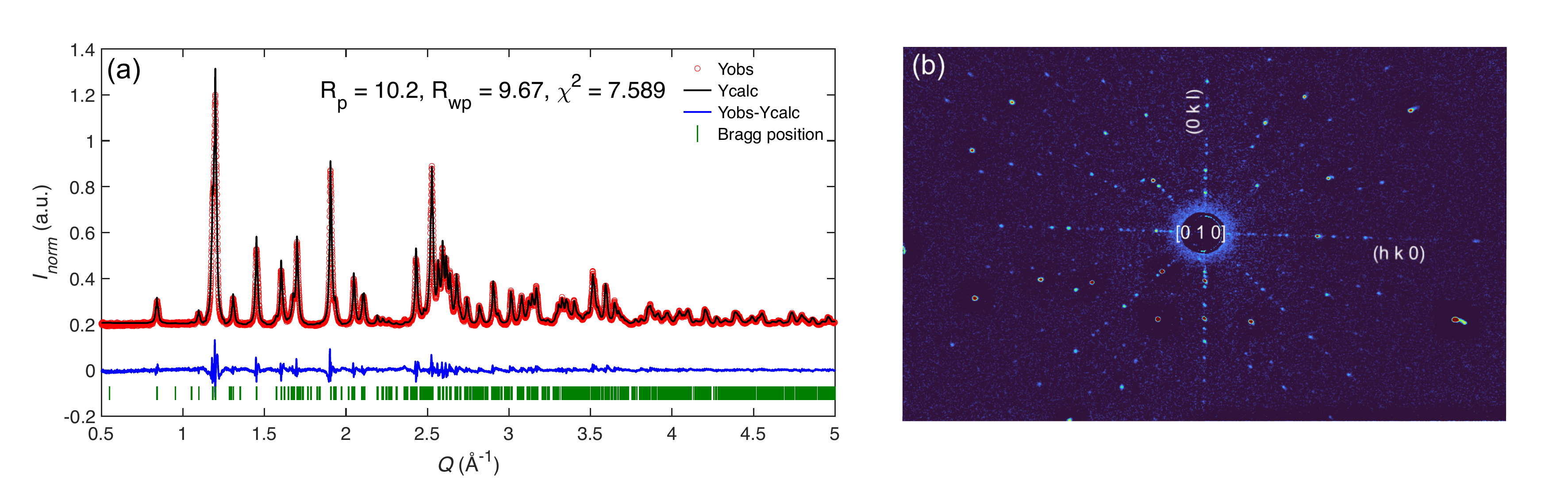}

\caption{\label{Sfig:Xray} (a) Rietveld refinement of the powder X-ray diffraction pattern collected on crushed single-crystal Henmilite, showing observed data ($Y_{\text{obs}}$), calculated profile ($Y_{\text{calc}}$), allowed Bragg peak positions, and the difference curve ($Y_{\text{obs}} - Y_{\text{calc}}$). (b) Single-crystal Laue diffraction pattern of Henmilite recorded with the incident X-ray beam aligned along the $b^*$ reciprocal direction. The horizontal row of reflections corresponds to the $a^*$--$c^*$ reciprocal plane (indexed as $(h\,0\,l)$), whereas the vertical array represents the $b^*$--$c^*$ plane ($(0\,k\,l)$ reflections). The diffraction data were collected using the ALSA automated platform.}
\end{figure}

\section{Transition and Spin Gap}

\begin{figure}[htbp]
    \centering
    \graphicspath{{figures_supplement/}}
    \includegraphics[width=0.9\linewidth]{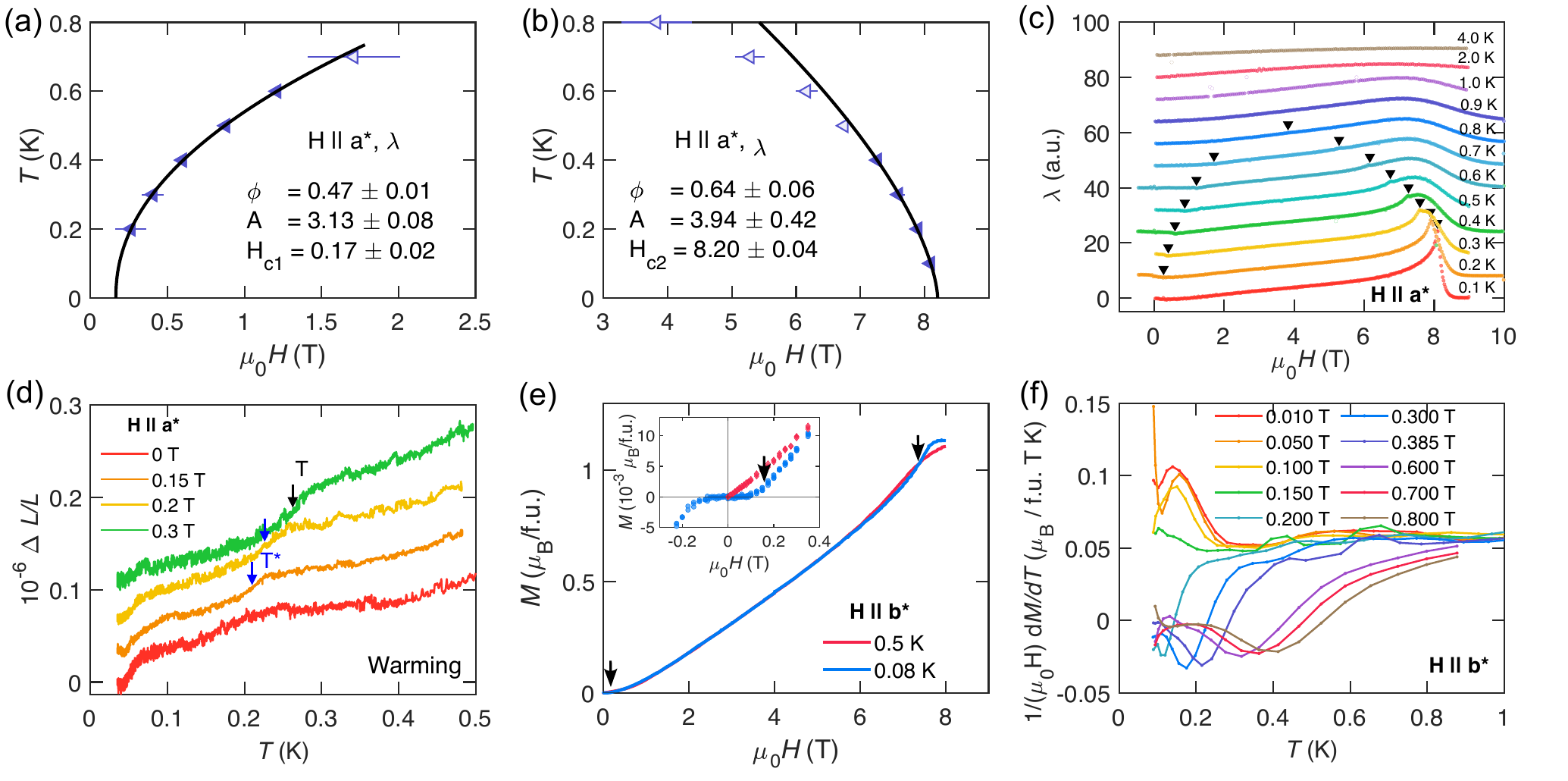}
\caption{\label{Sfig:2} Phase boundary tracking and scaling law fits obtained from the magnetostriction coefficient $\lambda$ for $H \parallel a^*$. The solid black lines represent fits to the power-law functions $H(T) = H_{\text{c1}} + A \cdot T^{1/\phi}$ for the lower boundary (left panel) and $H(T) = H_{\text{c2}} - A \cdot T^{1/\phi}$ for the upper boundary (right panel). Filled (dark) markers indicate data points included in the fitting procedure, while open (light) markers denote phase boundary points excluded from the fit. The resulting fit parameters ($\phi$, $A$, $H_{\text{c1}}$, and $H_{\text{c2}}$) are displayed within each panel. (c) Magnetostriction data, shifted vertically for clarity. Magnetic transitions have been marked with black triangles. (d) Temperature dependence of the thermal expansion at low fields. The curves are shifted vertically by $0.035$ units. Black arrows mark magnetic transition points (T) while blue arrow show T*. (e) Field dependence of magnetisation measured down to dilution temperatures. The inset shows an expanded view at low magnetic fields. Arrows represent the critical fields. 
(f) Temperature derivative of the magnetic susceptibility,
$\mathrm{d}(M/\mu_0H)/\mathrm{d}T$, for applied fields from
$\mu_0H=0.010$ to $0.800$~T.
The corresponding colour map is shown in Fig.~1f of the main text.
}
\end{figure}

The identification of a phase boundary in the $H$-$T$ phase diagram depends on the criterion used to define the transition point. 
This is important because different criteria can affect the extracted scaling exponent and critical field. Based on the phase boundary obtained from magnetostriction data for $H \parallel a^*$, presented in Fig.~\ref{Sfig:2}~(a-b), the lower critical field $H_{\mathrm{c1}}$ was determined by fitting the experimental data to the scaling relation $H(T) = H_{\mathrm{c1}} + A T^{1/\phi}$. 
The fit yields an extrapolated zero-temperature critical field of $\mu_0H_{\mathrm{c1}} = 0.17 \pm 0.02~\text{T}$ and a scaling exponent of $\phi = 0.47 \pm 0.01$. 
This power-law behaviour is characteristic of the phase-boundary scaling typically observed in quantum spin-gap systems undergoing field-induced phase transitions. It should be noted that the obtained lower-critical-field exponent differs from the value $\phi = 2/3$ expected for a pure three-dimensional BEC system, whereas the exponent at the upper critical field is consistent with $2/3$ within the error bars. 
Further discussion is provided in the section \hyperref[sec:BEC]{Bose--Einstein condensation (BEC) of triplons}. The magnetic transition temperatures and critical fields are marked by black arrows in the thermal expansion curves (Fig.~\ref{Sfig:2}d) and the field dependence of the magnetisation at 0.08~K (Fig.~\ref{Sfig:2}~e), respectively. The inset of Fig.~\ref{Sfig:2}~e shows an enlarged view of the low-field regime. 
Figure~\ref{Sfig:2}f shows
$\mathrm{d}(M/\mu_0H)/\mathrm{d}T$ at selected magnetic fields. The
field-induced transition is identified by a minimum, consistent
with the positive slope $\mathrm{d}T_c/\mathrm{d}H>0$ of the lower
phase boundary. Below $H_{c1}$, the positive maximum defines the
crossover scale $T^*$ associated with thermal activation across the
small spin gap.

The intrinsic zero-field spin gap $\Delta$ is directly related to this critical threshold via the Zeeman energy relation $\Delta = g\mu_{\mathrm{B}}~\mu_0H_{\mathrm{c1}}$. Due to the $\sim$10\% variation of the $g$ factor within the $bc$ plane \cite{Hayashi_JPSCP_2023}, where it ranges from $2.05$ to $2.25$, the gap was estimated using the average value $g = 2.15$. This gives $\Delta/k_{\mathrm{B}} \approx 0.245~\text{K}$, corresponding to $\Delta \approx 21~\mu\text{eV}$. Including both the variation of the $g$ factor and the uncertainty in $H_{\mathrm{c1}}$, we obtain $\Delta/k_{\mathrm{B}} = 0.25 \pm 0.08~\text{K}$, or equivalently $\Delta = 21 \pm 7~\mu\text{eV}$.

\subsection{Magnetic Entropy Estimation}

\begin{figure}[htbp]
    \centering
    \graphicspath{{figures_supplement/}}
    \includegraphics[width=0.6\linewidth]{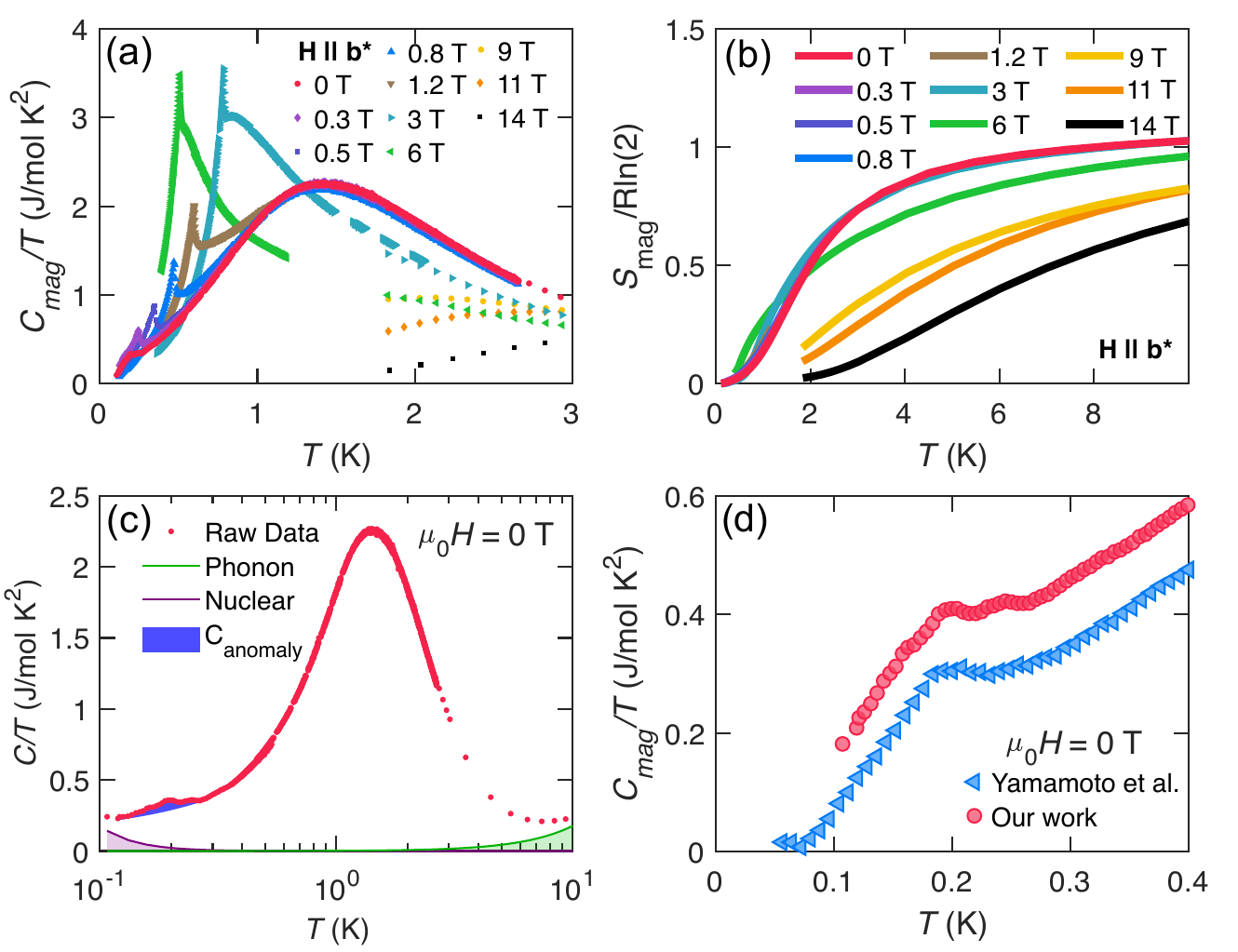}
    \caption{\label{Sfig:3}
    (a) Temperature dependence of the magnetic specific heat for $H\parallel b^*$ at selected magnetic fields. Sharp $\lambda$-shaped anomalies mark transitions into the field-induced ordered phase. The broad maximum centred near 1.5~K reflects the development of short-range magnetic correlations. Data collected in $^4$He, $^3$He, and dilution refrigerators have been combined.
    (b) Corresponding magnetic entropy for selected applied fields, obtained after subtraction of the phonon and nuclear Schottky contributions, for $H\parallel b^*$.
    (c) Zero-field raw specific heat plotted on a logarithmic temperature scale, together with the nuclear and phonon contributions. The shaded area marks the weak low-temperature anomaly. Its very small entropy and the $\mu$SR results are inconsistent with its interpretation as conventional full-volume magnetic long-range order.
    (d) Comparison of the magnetic specific heat, $C_{\mathrm{mag}}/T$, with those reported by Yamamoto \textit{et al.}~\cite{yamamoto2021quantum}. Our data have been shifted upward by $0.1~\mathrm{J\,mol^{-1}\,K^{-2}}$ for clarity. 
    }
\end{figure}

Specific heat measurements were performed across three temperature regimes: using the standard relaxation technique in a PPMS equipped with a helium-4 ($^4\text{He}$) cryostat, via a long-pulse method \citep{Scheie_JLTP_2018} with dual-slope data acquisition in a helium-3 ($^3\text{He}$) cryostat, and within a Triton dilution refrigerator. These datasets were merged to form a single, continuous temperature profile. Measurements were conducted along various crystallographic directions (not shown) and in applied magnetic fields up to 14~T (Fig.~\ref{Sfig:3}a and corresponding magnetic entropy in Fig.~\ref{Sfig:3}b).
To isolate the magnetic contribution, $C_{\mathrm{mag}}$, the phonon background, modelled as $C_{\mathrm{phonon}}=\beta T^3$ with $\beta=1.772\times10^{-3}~\mathrm{J\,mol^{-1}\,K^{-4}}$, and the low-temperature nuclear Schottky contribution were subtracted from the total specific heat, $C_{\mathrm{raw}}$ (Fig.~\ref{Sfig:3}c):
\[
C_{\mathrm{mag}}=C_{\mathrm{raw}}-C_{\mathrm{phonon}}-C_{\mathrm{nuc}}.
\]

Because measurements down to approximately 100~mK do not fully resolve the nuclear upturn, we used the nuclear Schottky coefficient reported by Yamamoto \textit{et al.}~\cite{yamamoto2021quantum}, whose measurements extended to lower temperatures.

Figure~\ref{Sfig:3}d compares the resulting low-temperature magnetic specific heat with the previously published data of Yamamoto \textit{et al.}~\cite{yamamoto2021quantum}. The weak low-temperature anomaly appears in both datasets with similar temperature dependence, demonstrating that it is reproducible rather than a measurement artefact or impurity. Taken together with the magnetisation, magnetoelastic, and $\mu$SR results, we identify this feature as the thermodynamic signature of the gap-controlled crossover at $T^*$ into the quantum-disordered regime, rather than a phase transition into conventional bulk magnetic long-range order. The associated entropy is only a minute fraction of the total spin entropy, 
$\Delta S\lesssim0.15\%\,R\ln2$ per Cu (see Fig.~1d of the main text).


\section{Neutron Diffraction}

In a field-induced triplon-condensation scenario, the ordered phase inside the dome is expected to carry a transverse antiferromagnetic component associated with phase coherence of the bosonic quasiparticles. Elastic neutron diffraction is therefore the natural microscopic probe of the field-induced order. In Henmilite, however, such an experiment is intrinsically challenging: the natural material contains boron and hydrogen, leading respectively to neutron absorption and a large incoherent background, while the available natural single crystals are small.

To maximise the magnetic signal relative to absorption and background, we selected the largest available Henmilite single crystal for neutron diffraction. An ideal experiment would require a significantly larger crystal, of order $150$~mg, whereas the largest crystal available for the present study had a mass of $76$~mg. Several neutron diffraction experiments were attempted on this crystal using different instruments and configurations, including D10 at ILL, ZEBRA at PSI, and WISH at the ISIS Neutron and Muon source. These measurements did not yield a resolvable magnetic signal above the experimental background. Details of these measurements are summarised below.

At D10 (Institut Laue--Langevin), we used the largest Henmilite single crystal inside an orange cryostat with dilution insert; the scattering plane was  $a^*-c^*$ \citep{ILL_DATA_5_41_1274}. We utilised an 8~mm radial collimator before the sample, a PG analyser, and a vertically focused beam at the sample position. The experimental paths consisted of continuous $H$-, $L$-, and diagonal linear scans mapping the $(H, 0, L)$ reciprocal space region bounded between $(-2, 0, 0)$ and $(0, 0, -2)$. No magnetic signal was found. At Zebra (Paul Scherrer Institute \citep{PSI}), we continued the measurements on the same sample with a more precise orientation, utilising a dilution cryomagnet with a maximum magnetic field of 3~T ($T_{\text{base}} \sim 50$~mK). We used a Ge-220 monochromator with a $\pm14^\circ$ vertical opening, along with both 2D and 1D detectors. The measurement paths covered the $(H, 0, L)$ plane as well as the $(H, 1/2, L)$ plane at $\mu_0H = 0$ and 3~T. At WISH (ISIS Neutron and Muon source, UK \citep{WISH}), a powder sample of mass $0.95$~g was distributed in thin layers between four sheets of copper foil and enclosed in a double-walled copper can \citep{ISIS_RB2420429_WISH_1}. This layered geometry reduced the effective neutron path length through the sample, thereby mitigating neutron absorption by $^{10}$B at natural isotopic abundance. A separate single crystal experiment \citep{ISIS_RB2400089_WISH_2} at $\mu_0H = 0$~T and maximum vertical fields up to 6~T in a$^*$-c$^*$ scattering plane was conducted as well. We collected data for around 7 hours at each $(H, T)$ configuration to gather more statistics. Unfortunately, none of these experiments revealed any magnetic Bragg peaks inside the dome phase.

We therefore performed a further experiment on the cold-neutron
triple-axis spectrometer THALES (Institut Laue--Langevin) \citep{ILL_DATA_TEST_3445}. The crystal
was mounted with the $(H\,0\,L)$ plane horizontal in the vertical-field
cryomagnet. Its in-plane orientation was established from $\omega$ scans
through the strong, non-collinear nuclear reflections $(0\,0\,1)$ and
$(2\,0\,0)$, which span the $a^*$--$c^*$ scattering plane. These scans
were used to determine the sample-angle offset and were repeated after
changes of the neutron wave vector and beam-defining optics. The broad
vertical acceptance of the TAS configuration made small out-of-plane
misorientations non-limiting for the present measurement.

\begin{figure}[htbp]
    \centering
    \graphicspath{{figures_supplement/}}
    \includegraphics[width=1\linewidth]{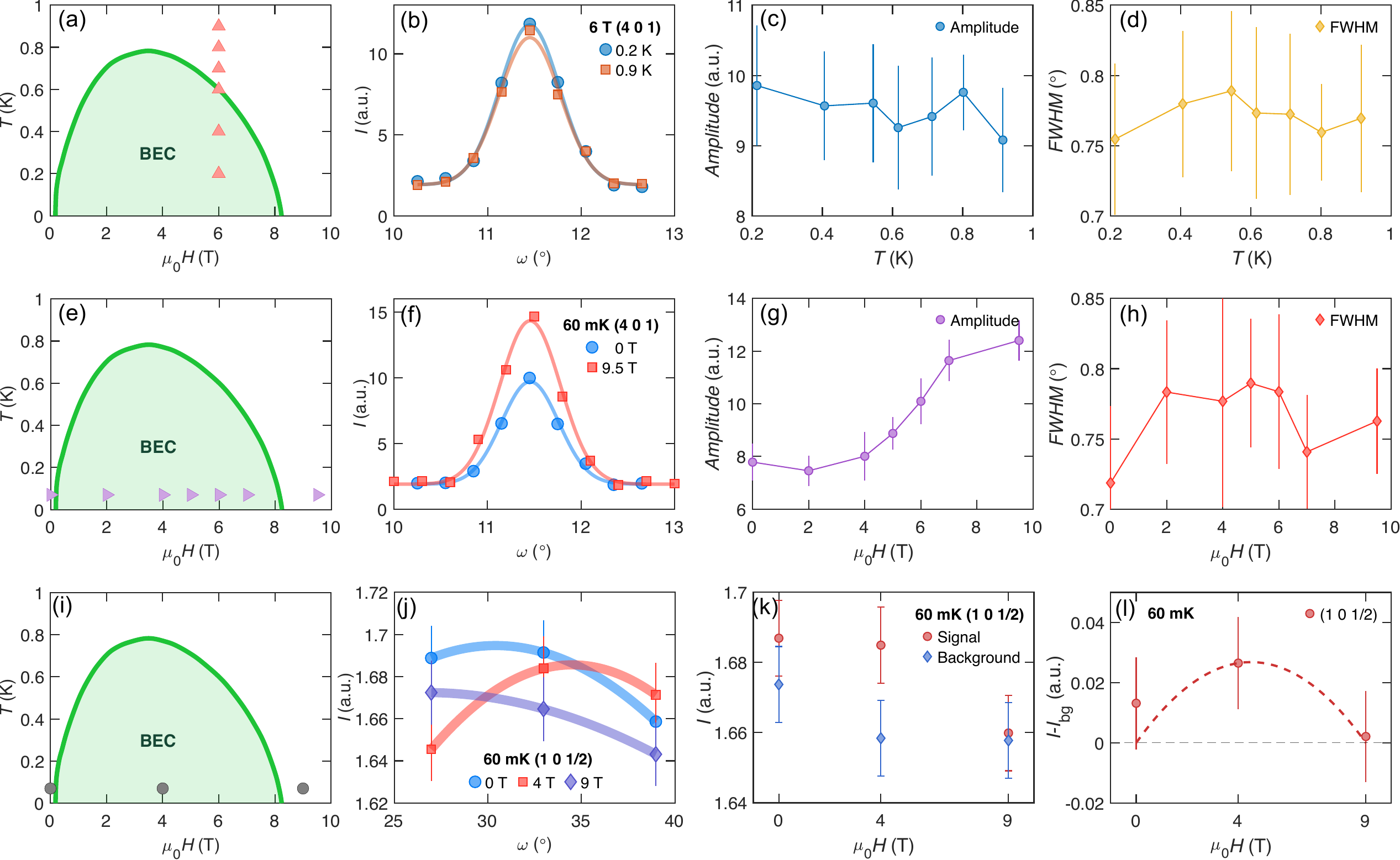}
\caption{\label{Sfig:4} 
(a,e) Schematic $H$--$T$ phase diagrams showing the data points used in the
measurement of the peak $(4\,0\,1)$.
(b--d) Temperature dependence and (f--h) field dependence of the
$(4~0~1)$ nuclear reflection, together with the corresponding
Gaussian-fit parameters. The peak centre and background were fixed to
their mean values, which were insensitive to temperature and field
within uncertainty.
(i) Schematic $H$--$T$ phase diagram showing the fields used in the
search at the candidate antiferromagnetic position
$(1\,0\,\tfrac{1}{2})$.
(j) $\omega$ scans covering this position at selected fields.
(k) Central-point intensity and local background estimated from the two
symmetrically placed side points at $\mu_0H=0$, $4$, and $9$~T.
(l) Difference between the central-point intensity and the local
background estimate. 
The horizontal dashed line marks zero; the red dashed curve is a 
schematic guide showing the field dependence expected for a signal 
confined to the ordered dome and is not a fit to the data. 
No statistically significant field-induced Bragg peak is resolved.
}
\end{figure}

Because the expected magnetic signal was weak, a substantial part of the
beam time was devoted to optimising the elastic configuration for the
best signal-to-background ratio. We tested several values of
$k_{\mathrm{i}}=k_{\mathrm{f}}$, including higher-flux long-wavelength
settings, both filtered and unfiltered configurations, different primary
and secondary slit openings, and alternative post-sample collimation
geometries. The nuclear reflections $(0\,0\,1)$ and $(2\,0\,0)$ were
used to compare integrated intensity, rocking width, and background.
This procedure identified
$k_{\mathrm{i}}=k_{\mathrm{f}}=1.45~\text{\AA}^{-1}$ as the best
compromise between flux and background. The final elastic configuration
used a horizontally flat PG monochromator and a conventional $60'$
collimator after the sample, rather than the radial collimator, to improve
the in-plane momentum resolution.

Following this optimisation, the intensity of the nuclear Bragg reflection $(4~0~1)$ was tracked systematically as a function of external magnetic field at the base temperature of the $^3$He/$^4$He dilution refrigerator ($\sim 60$~mK).
The fitted peak amplitude increases systematically with field, whereas
the peak position and width remain unchanged within uncertainty,
Fig.~\ref{Sfig:4}(e--h). After subtraction of the zero-field nuclear
baseline, the field-induced intensity follows
$[M_{\mathrm{QMC}}(H)]^2$ after application of a single scale factor,
consistent with a field-induced $\mathbf{k}=0$ magnetic contribution,
as shown in Fig.~3b of the main text.
Fig.~\ref{Sfig:4} summarises the corresponding  temperature and field dependence.

In addition, we searched for the transverse antiferromagnetic Bragg scattering expected inside the field-induced ordered dome. Guided by the model-motivated propagation vector $\mathbf{k}=(0,0,\tfrac{1}{2})$, we selected $(1,0,\tfrac{1}{2})$ as the strongest accessible candidate reflection for the available sample orientation and instrument geometry. However, neither the actual propagation vector nor the direction of the transverse ordered moment—and hence the relative intensities of the magnetic reflections—is known experimentally. The selected reflection should therefore be regarded as one plausible search position rather than a uniquely expected magnetic peak.

The nominal $\omega$ position of $(1,0,\tfrac{1}{2})$ was calculated from the sample alignment established using the nuclear reflections $(0,0,1)$ and $(2,0,0)$. Before the extended-counting measurements (30 minutes per point), conventional full $\omega$ scans covering the candidate position were recorded at $\mu_0H=0$ and $4$~T. No resolved peak profile was observed in either scan. To maximise the counting statistics at the nominal peak position, subsequent measurements used a three-point scheme at $\mu_0H=0$, $4$, and $9$~T, representing fields below, inside, and above the ordered dome, respectively. The central point was counted for twice as long as each of the two symmetrically placed side points. The two side-point counts were combined, so that their total counting time matched that of the central point, and were used to estimate the local background.

The resulting central-point and background intensities, together with their difference, are shown in Figs.~\ref{Sfig:4}(k,l). The data were normalised to the incident-beam monitor and acquisition time, and the uncertainties of the difference were propagated from the counting statistics of both the central and side-point measurements. At $\mu_0H=4$~T, the central-minus-background difference is positive and its plotted counting-statistical error bar does not include zero, whereas the corresponding differences at $\mu_0H=0$ and $9$~T are consistent with zero. However, the three-point measurement does not resolve a peak profile, and the local background is constrained by only two side points. Consequently, uncertainties associated with the background shape and with the unknown position and width of a possible peak are not included in the plotted counting-statistical error bar. The difference shown in Fig.~\ref{Sfig:4}(l) should therefore be regarded as a local contrast estimate rather than a fitted or integrated magnetic Bragg intensity. These data do not provide sufficiently robust evidence for a field-induced antiferromagnetic reflection and are not used to determine the magnetic propagation vector or staggered moment.


\section{MUON SPIN RELAXATION}

A $\mu$SR experiment \citep{ISIS_RB2310640_HIFI} was performed on a powder sample at the HiFi beamline~\cite{HiFi} at the ISIS Neutron and Muon Source (UK) using fully spin-polarised positive muons ($\mu^{+}$). Measurements were carried out in zero-field (ZF), longitudinal-field (LF), and transverse-field (TF) configurations.

In $\mu$SR experiments, implanted muons probe the sample-averaged local magnetic fields arising from different sites within the crystallographic and magnetic unit cell. Henmilite crystallises with 12 distinct hydroxyl (–OH) groups and two inequivalent Cu sites in a unit cell, both of which are expected to provide favourable stopping sites for implanted muons. For the experiment, a powder sample was prepared by crushing and grinding single crystals of natural mineral Henmilite, followed by sieving through a $50\,\mu$m mesh to obtain finer grains. The resulting powder was pressed into a pellet of a diameter 18 mm and of a total mass $\sim 0.6$ g under an applied uniaxial load of approximately 320 kg. A small amount of diluted GE varnish (in toluene and ethanol) was applied to both faces of the pellet to improve its mechanical stability and thermal contact between grains to help with temperature stabilisation. 
The pellet was mounted on a square silver plate using a thin layer of
Apiezon N grease between the pellet and the plate to provide thermal
contact and mechanical stability.

Good thermalisation of the powder pellet is supported by two
independent observations. First, the transition temperatures determined
from the finite-field relaxation rate $\lambda(T)$
(Fig.~2d of the main text) agree with the phase boundaries obtained
from single-crystal thermodynamic and magnetoelastic measurements
(see stars in Fig.~1e of the main text). Second, specific-heat measurements on the
pressed-pellet material reproduced the transition temperatures observed
in the single crystals. 

To ensure that muons predominantly stop within the sample, the need for additional silver attenuation was carefully evaluated. A single layer of Ag foil (thickness $12.5\,\mu$m) was placed over the sample, serving to secure the pellet and to optimise the muon stopping profile by reducing undesired implantation into the silver holder giving background. At the HiFi instrument, divergent slits (slit-16 configuration) were used, corresponding to a muon flux of approximately $40\,\mathrm{MEvents\,hr^{-1}}$ at the sample position. For TF measurements, a transverse magnetic field of 2 mT (TF20) was applied perpendicular to the initial muon spin polarisation, and the temperature dependence of the asymmetry was recorded upon warming from 50 mK to 4 K. ZF measurements were performed over the same temperature range under active magnetic field compensation at the sample position, even for Earth's magnetic field. For data analysis, the detector efficiency ratio $\alpha$ was determined from a weak transverse field TF20 measurement at 60 K by fitting the muon precession signals in the forward and backward detector groups, yielding $\alpha = 1.13119$. The value of $\alpha$ was fixed for all subsequent analyses. The data analysis was performed in Mantid~\cite{Mantid}.

\begin{figure}[htbp]
    \centering
    \graphicspath{{figures_supplement/}}
    \includegraphics[width=1\linewidth]{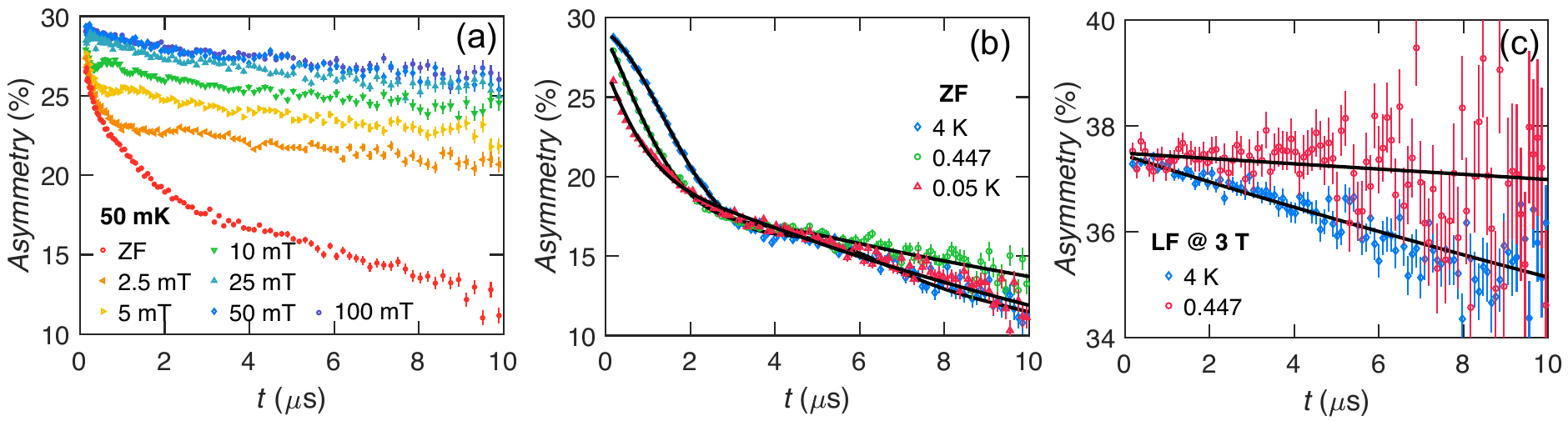}
    \caption{\label{Sfig:Musr_spectra} (a) Field dependence of the $\mu$SR asymmetry at $T$ = 50 mK measured in longitudinal field (LF) (obtained from WiMDA software \citep{wimda} with variable binning size). (b) zero-field (ZF) and (c) LF at $\mu_0H$ = 3 T spectra at $T$ = 0.05 K, 0.447 K and 4 K, together with fits.}
\end{figure}

Figure~\ref{Sfig:Musr_spectra} shows the $\mu$SR asymmetry measured at $T$ = 50 mK under applied longitudinal fields (LF). The relaxation at low fields is dominated by quasi-static local magnetic fields arising primarily from weakly coupled nuclear moments, which can be progressively decoupled by the application of a longitudinal field, but there is still a contribution from electronic moments. In the present data, significant decoupling is observed for fields of order 25 mT, above which the spectra become weakly field-dependent, indicating suppression of the static local-field contribution, including that associated with $\mu$–OH complexes. At higher fields, the relaxation is dominated by residual dynamic (electronic) fluctuations. The relatively small decoupling field ($\sim 25$ mT) suggests that the internal static fields were weak, consistent with the absence of long-range magnetic order at the base temperature.

In Henmilite, the proximity of the muon to the proton in an –OH group can lead to the formation of an entangled $\mu$–OH complex, characterised by a well-defined muon polarisation function [Eq.~(\ref{P_OH})], analogous to the F–$\mu$–F state~\cite{Lancaster_PRL_2007}. This contribution is observed as a weak oscillatory feature (hump) in the asymmetry around $t \sim 4\,\mu$s in the ZF data (Fig.~\ref{Sfig:Musr_spectra}), which shifts to shorter time window with increasing longitudinal field. To capture this hump, a $\mu$-H polarisation function was used \cite{LORD2000495}:

\begin{equation}
\label{P_OH}
    P_{\mathrm{OH}}(t) = \frac{1}{6} \left[1 + \cos\!\left(\omega t + \phi\right) + 2 \cos\!\left(\frac{\omega t}{2} + \phi\right)
    + 2 \cos\!\left(\frac{3\omega t}{2} + \phi\right)
    \right]
\end{equation}

\noindent where $\phi$ is a phase shift at t= 0 which was fixed to zero throughout the analysis, and $\omega$ is the oscillation frequency related to the $\mu$-OH separation distance. Such polarisation functions have been found to fit numerous zero-field muon relaxation spectra for hydroxyl-cuprate magnetic minerals\cite{Fak_PRL_2012, Tustain_NPJQM_2020, Berlie_PRB_2022}. In our sample, spin fluctuations are too rapid to couple to the muon in the temperature range 300 K$\leq$ T $\leq$4 K, therefore the relaxation spectra in this temperature window appear temperature-independent and can be fitted by a complete function:

\begin{equation}
    \label{eqn:2}
    A(t) = A_1*P_{OH}*e^{-\frac{(\sigma t)^2}{2}} + A_2*e^{-\lambda_2 t}
\end{equation}

The slowly relaxing baseline term ($A_2 e^{-\lambda_2 t}$) accounts for the contribution of muons implanted outside the sample, primarily in the silver sample holder. 
The first term in eqn.\ref{eqn:2} arises due to a large number of crystallographically distinct hydroxyl groups in Henmilite, resulting in a distribution of local muon environments that will each have a slightly different oscillation frequency, broadening the oscillations and modelled by a Gaussian decay term, $\sigma$ (see eqn. \ref{eqn:2}).

For T $\leq$ 4 K, the electronic fluctuations are sufficiently slow to couple to the muons, resulting in an additional depolarisation pathway. This additional relaxation was modelled by a Lorentzian depolarisation term, $\lambda_1$. The revised equation becomes the following:

\begin{equation}
    \label{eqn:3}
    A(t) = A_1*P_{OH}*e^{\frac{-(\sigma t)^2}{2}}*e^{-\lambda_1 t} + A_2*e^{-\lambda_2 t}
\end{equation}

The zero-field (ZF) data were fitted by fixing the $P_{\mathrm{OH}}$ parameters and the Gaussian relaxation rate $\sigma$ to their high-temperature values determined at $T = 60~\mathrm{K}$. Within the $P_{\mathrm{OH}}$ polarization function, the angular frequency $\omega$ is given by $2\pi\nu_{\mathrm{D}}$, where the oscillation frequency $\nu_{\mathrm{D}}$ was held constant at $0.1113~\mathrm{MHz}$ and $\sigma$ was fixed at $0.26~\mu\mathrm{s}^{-1}$. Detailed results are discussed in the main text (refer to Fig. 2b). Muon spectra for $\mu_0H \ge$ 25 mT were fitted by just a Lorentzian term and background: $A(t) = A*e^{-\lambda_{LF}~t} + bg$.

Transverse-field (TF) $\mu$SR measurements provide information on the volume fraction of static (frozen) and dynamic spins~\cite{Tustain_NPJQM_2020}. The weak-TF measurement was performed by applying a small magnetic field of 20 G perpendicular to the initial muon spin polarisation. For the analysis of the TF spectra, the following strategy was adopted. In fitting the spectra for $T \leq 4$ K, the temperature-independent nuclear contribution was fixed to the values obtained at higher temperatures, and only the additional electronic relaxation and the baseline contribution were allowed to vary. All fits were performed using the Mantid software package~\cite{Mantid}. The following equation represents the fitting function:

\begin{equation}
\label{TF20}
    A_{TF} = \left[A_d + A_p e^{-\lambda t} + A_b e^{-\sigma^2 t^2}\right]
    \cos\!\left( 2\pi \nu t + \phi \right) + B
\end{equation}

Here, $A_d$ represents the non-relaxing component associated with strong spin fluctuations, $A_p$ is the Lorentzian relaxation component arising from electronic spin dynamics, and $A_b$ accounts for the Gaussian relaxation due to the silver sample holder. $\lambda$ and $\sigma$ denote the corresponding relaxation rates from Lorentzian and Gaussian relaxations, respectively, $\nu$ is the muon precession frequency, $\phi$ is the initial phase, and $B$ is a constant background.

\begin{figure}[ht]
    \centering
    \graphicspath{{figures_supplement/}}
    \includegraphics[width=0.7\linewidth]{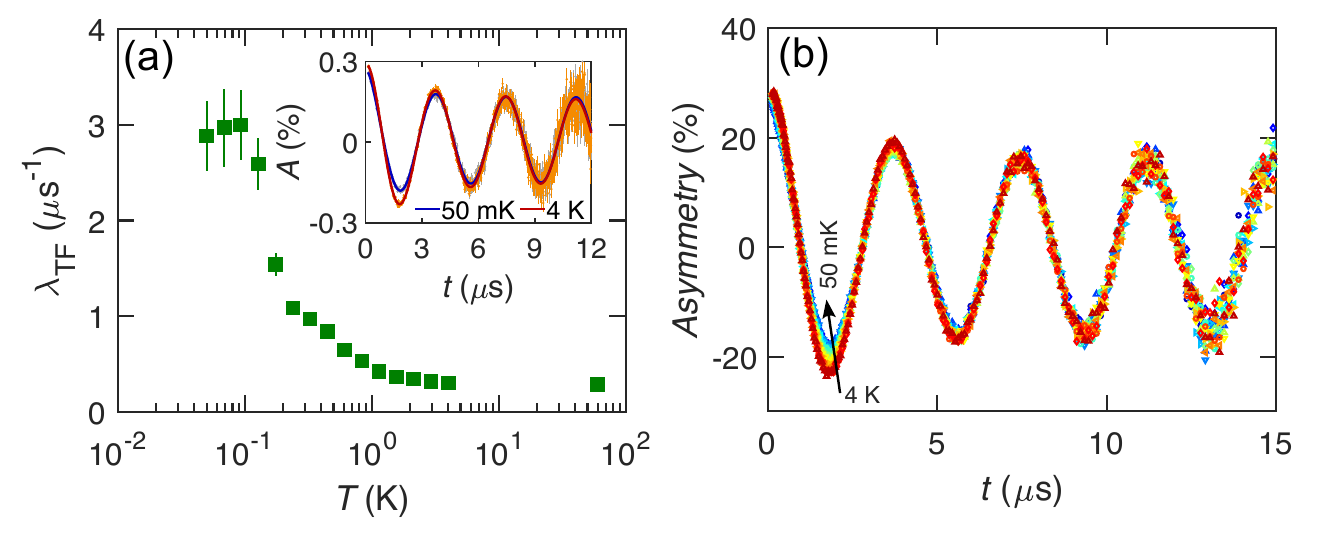}
    \caption{\label{Sfig:FrozenFraction} 
    (a) Temperature dependence of the transverse-field relaxation rate
    $\lambda_{\mathrm{TF}}$ obtained from the TF20 measurements; the inset
    shows representative spectra and fits. 
    (b) TF asymmetry spectra over
    the full measured temperature range. The frozen fraction derived from
    the fitted amplitudes is shown in Fig.~2c of the main text.
    }
\end{figure}

To systematically isolate the effect of spin fluctuations at low fields, the parameters $A_b$, $\sigma$, $\phi$, and $\nu$ were fixed to their values obtained from the 60 K dataset, where all parameters in Eq.~(\ref{TF20}) were allowed to vary freely. The value of $A_d$ was fixed to its corresponding value at 50 mK, where it exhibits a plateau. In fitting the temperature dependence of the TF data, only $A_p$, $\lambda_{\mathrm{TF}}$, and $B$ were allowed to vary. A frozen fraction was defined from $A_p$ (see Eq.~(\ref{FrozenFraction}) and ref~\cite{Tustain_NPJQM_2020}). The results of the TF20 measurements are shown in Fig.~\ref{Sfig:FrozenFraction}. The temperature dependence of $\lambda_{\mathrm{TF}}$ is consistent with that obtained from ZF measurements (see main text Fig. 2), supporting the validity of the fitting procedure.

\begin{equation}
\label{FrozenFraction}
\mathrm{FF} = 100 \left[ 1 - \frac{(A_d + A_p)_T}{(A_d + A_p)_{60\,\mathrm{K}}}
\right]
\end{equation}

\section{DFT and QMC}

To model Henmilite's electronic and magnetic behaviour, density functional theory (DFT) and quantum Monte Carlo (QMC) methods were employed. In the DFT calculations, the electronic band structure was evaluated together with the leading magnetic exchange interactions (Eq.~\ref{Eq.:HeisenbergHam}).
Ab initio calculations were performed within the Vienna ab initio simulation package (VASP)~{\cite{Kresse_VASP_r96,Kresse_VASP_r99}} using  projector-augmented-wave (PAW) method pseudo-potentials and generalised gradient approximation (GGA) of Perdew-Burke-Ernzerhof
(PBE)~\cite{PBE_r96}. The energy cut-off for the plane waves was 450 eV and 3$\times$4$\times$6 $\Gamma$-centered mesh was considered employing the tetrahedron method Brillouin zone integrations with Blöchl corrections~\cite{Bloechl_corr_tetrahedron}. In addition, the van der Waals corrections of DFT-D3 method (IVDW=11)~\cite{DFT_D3_IVDW11} were included. The calculated Cu magnetic moments $|\mu(\mathrm{Cu})|=0.58 \mu_{\mathrm{B}}$ correspond well to the $S=1/2$ character of the compound.
To obtain the magnetic pair exchange interactions, the maximally localised Wannier functions~\cite{MaxlocWF} as implemented in the Wannier90 package~\cite{Wannier90} were employed, considering the Cu 3d- and O 2p- states, where the exchange interactions were evaluated based on the Wannier states within the TB2J package~\cite{TB2J} on the 12$\times$16$\times$24 k-mesh using the Liechtenstein-Katsnelson-Antropov-Gubanov approach~\cite{LKAG}.
Both collinear calculations and non-collinear ones with spin-orbit interactions were performed, while the difference was negligible. Therefore, only the results for the collinear calculations are presented.

The three leading exchange couplings, $J_{\mathrm{leg}}$, $J_{\mathrm{rung}}$, and $J_{\mathrm{inter}}$, show an approximately systematic variation with the Cu--Cu separation, highlighting the sensitivity of the magnetic interactions to the underlying crystal geometry. In our DFT calculations, no bands cross the Fermi level, in contrast to the previous report~\cite{yamamoto2021quantum}, thereby supporting the insulating nature of the mineral.

\begin{figure}[ht]
    \centering
    \graphicspath{{figures_supplement/}}
    \includegraphics[width=0.9\linewidth]{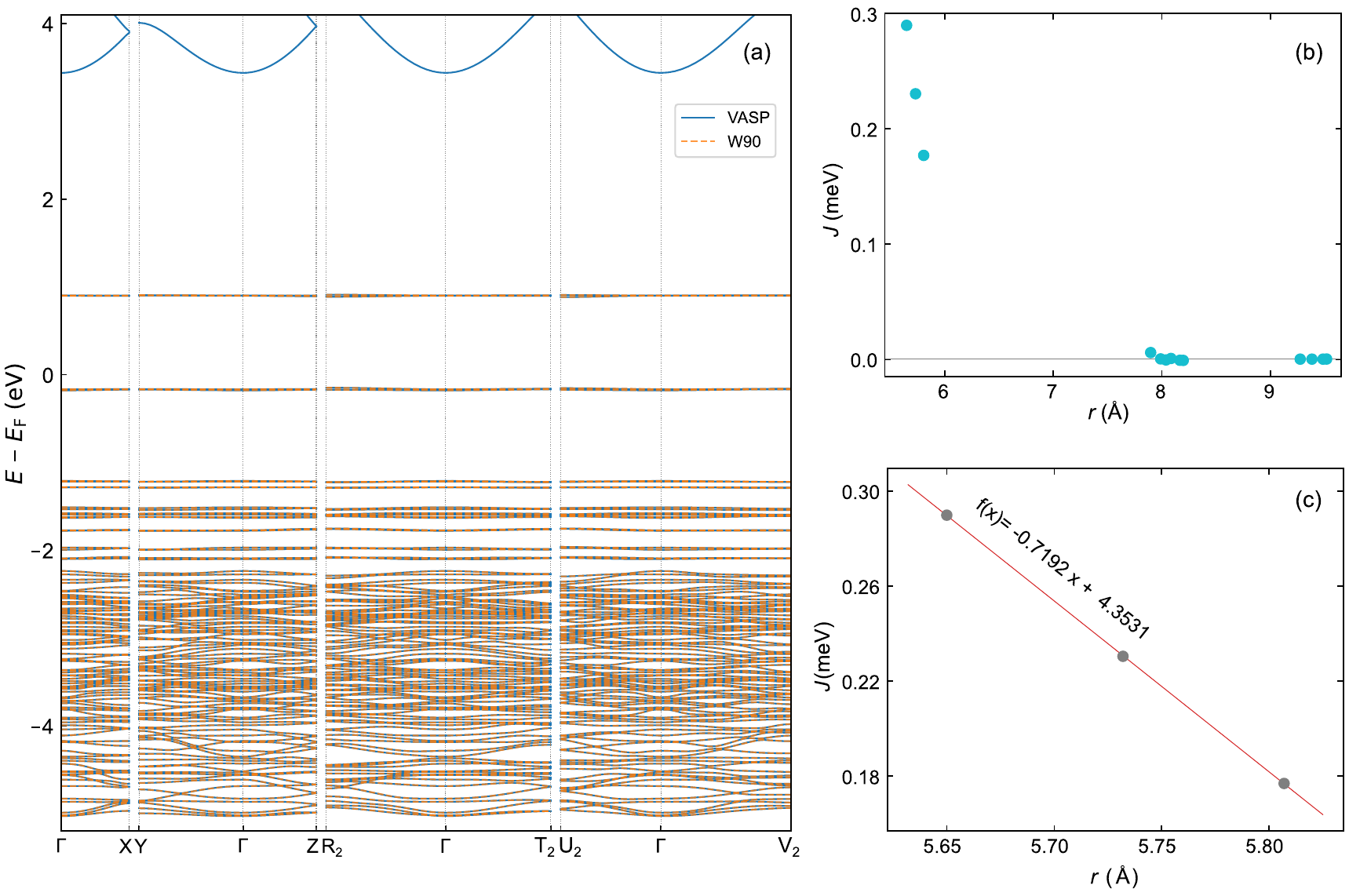}
  \caption{\label{Sfig:Dominik} (a) Band structure of Henmilite, comparison of (solid lines) DFT results with (dashed lines) wannierized band structure. (b)  Exchange coupling $J_{ij}$ as a function of the Cu-Cu distance. (c) Magnified view of the previous dependence for three leading exchange couplings, $J_{\mathrm{leg}}$, $J_{\mathrm{rung}}$, and $J_{\mathrm{inter}}$, in Henmilite interpolated by linear dependence.}
\end{figure}

Guided by the DFT exchange topology, we restrict the QMC model to the three leading couplings, $J_{\mathrm{leg}}$, $J_{\mathrm{rung}}$, and $J_{\mathrm{inter}}$, which form a two-dimensional coupled-ladder lattice. QMC simulations for this model were performed using the loop~\cite{Todo_PRL_2001} and dirloop$\_$sse~\cite{Alet_PRE_2005} algorithms of the ALPS simulation package~\cite{Albuquerque_ICM_2007} on $L\times L$ finite lattices with $L$ up
to 24 and periodic boundary conditions. The critical temperatures of
field-induced magnetic ordering were determined as BKT temperatures for
the 2D spin Hamiltonian using the numerical procedure described in Ref~\cite{Harada_PRB_1997}. The good agreement between such 2D values
and the experimental phase boundary suggests the negligible role of the
interlayer couplings $J_{\perp1}$ and $J_{\perp2}$. Ordered magnetic moments in Fig. 4a (of main text) were determined by the $1/L$ extrapolation of the staggered magnetisation calculated at $T/J_{\rm rung}=0.02$ for different values
of $L$.

In Table~\ref{tab:DFT_QMC} we compare exchange couplings extracted from DFT and obtained from the QMC fits to the experimental data. Although the individual exchange hierarchy differs between the different approaches, the total exchange scale is very similar in all cases. This indicates that DFT captures the relevant few-kelvin magnetic energy scale, but the detailed low-energy hierarchy must be constrained by comparison between experiment and QMC.

\begin{table}[htbp]
\centering
\renewcommand{\arraystretch}{1.4}
\setlength{\tabcolsep}{12pt}
\begin{tabular}{lrrrr}
\textbf{Interaction} & \textbf{$d$ [\AA]} & \textbf{DFT} & \textbf{DFT} \cite{yamamoto2021quantum} & \textbf{QMC} \\ 
\midrule
$J_{\mathrm{leg}}$  & 5.650 & 3.4 & 3.6 & 3.0 \\
$J_{\mathrm{rung}}$  & 5.732 & 2.7 & 3.5 & 3.8 \\
$J_{\mathrm{inter}}$  & 5.807 & 2.1 & 0.9 & 1.3 \\
\midrule
$\sum J$             & --    & 8.2 & 8.0 & 8.1 \\
\end{tabular}
\caption{\label{tab:DFT_QMC}
Comparison of the leading exchange interactions obtained from the present DFT calculations, the previous DFT analysis of Yamamoto \textit{et al.}~\cite{yamamoto2021quantum}, and QMC fits to the spin-ladder model. Exchange constants are in Kelvin.}
\end{table}

\begin{equation}
    H_{\mathrm{ex}} =  \sum J_{ij}  \mathbf{e}_{i}  \mathbf{e}_{j},
    \label{Eq.:HeisenbergHam}
\end{equation}

The close agreement of the summed three leading couplings in Table~\ref{tab:DFT_QMC} shows that all approaches give a consistent overall magnetic energy scale. However, the redistribution of this exchange scale among leg, rung, and interladder couplings is method-dependent. This is important because the distinction between a weakly gapped coupled ladder and a zero-field ordered state is controlled not only by the bare exchange scale, but also by quantum fluctuations. Consequently, DFT alone is insufficient to identify the field-induced BEC regime or to determine on which side of the coupled-ladder ordering boundary Henmilite lies. For this reason, the effective ladder parameters used in the main text are fixed by QMC comparison with experiment, while DFT is used as an independent check of the exchange topology and energy scale.

\section{Bose--Einstein condensation (BEC) of triplons}
\label{sec:BEC}

The field-induced dome in Henmilite can be discussed within the standard triplon-condensation framework used for gapped quantum magnets~\cite{Zapf_RMP_2014}. In this description, the magnetic field tunes the chemical potential of bosonic triplet quasiparticles, and the ordered phase corresponds to transverse antiferromagnetic order with a coherent phase. Strictly speaking, an exact BEC description requires conservation of the boson number, which in spin language corresponds to an approximate U(1) symmetry of the magnetic Hamiltonian. In real materials this symmetry may be weakly broken, because anisotropic exchange, dipolar interactions, and spin-orbit-induced terms weakly break U(1) symmetry. Nevertheless, when these terms are small compared with the relevant exchange and thermal energy scales, the BEC framework remains an appropriate effective description of the field-induced ordered phase.

In triplon-condensation systems, the quasiparticles behave as hard-core bosons, reflecting the constraint that the underlying spin-$1/2$ Hilbert space allows only a limited local triplet density. This gives rise to a dome-shaped phase boundary in the temperature-field phase diagram. In an ideal particle-hole symmetric case the dome would be approximately symmetric about its midpoint in field. Pronounced asymmetry, however, has been reported in several materials, including \ce{Ba3Mn2O8}~\cite{Samulon_PRL_2009} and \ce{NiCl2.4SC(NH2)2} (DTN)~\cite{Kohama_PRL_2011_DTN}, where the phase boundary is tilted towards $H_{\mathrm{c1}}$. In this interpretation, the asymmetry reflects different effective masses of the low-energy bosonic quasiparticles near the lower and upper critical fields. In analogy with DTN, this can arise from quantum-fluctuation-induced mass renormalisation near $H_{\mathrm{c1}}$, whereas the excitations near $H_{\mathrm{c2}}$ remain closer to the bare high-field limit~\cite{Kohama_PRL_2011_DTN}.

Departures from the ideal BEC picture need not be limited to the
asymmetry of the outer phase boundary. In the ultralow-field dimer
magnet \ce{Yb2Si2O7}, thermodynamic and neutron measurements distinguish
low- and high-field regimes within the ordered dome
\cite{Hester_PRL_2019_Yb2Si2O7}. One theoretical description attributes
this internal structure to an anisotropy-driven zero-temperature phase
transition \cite{Flynn_PRL_2021_Yb2Si2O7}, whereas a later analysis
suggests that, at experimentally relevant temperatures, much of the
distinction may instead arise from a crossover associated with nonlinear
magnetisation \cite{Feng_PRB_2023_Yb2Si2O7}. Within the resolution of
our thermodynamic, magnetoelastic, and $\mu$SR measurements, no
reproducible additional phase boundary is resolved inside the Henmilite
dome, although weaker internal crossovers cannot be excluded.

Close to a quantum critical point, the phase boundary is expected to follow the scaling relation
$T_{\mathrm{c}} \propto (H-H_{\mathrm{c1}})^{\phi}$,
where $\phi$ is the critical exponent associated with the universality class of the transition. For a three-dimensional BEC of triplons, $\phi = 2/3$ is expected~\cite{Zapf_RMP_2014}. In Henmilite, the upper critical boundary gives $\phi \simeq 0.64$, consistent with this value. In contrast, the lower critical boundary gives $\phi \simeq 0.47$. We do not interpret this deviation as evidence for a different universality class. Rather, it likely reflects that the accessible temperature window does not reach the true asymptotic critical regime, or that crossover effects associated with anisotropy, reduced dimensionality, or proximity to the coupled-ladder instability remain relevant near $H_{\mathrm{c1}}$. The extracted exponent is also sensitive to the fitting window; therefore, the fits were tested by varying the number of included points and the fitted temperature range. This procedure yielded stable estimates of the critical fields used in the main text.

The pronounced asymmetry of the Henmilite dome can be quantified using the same effective-mass estimate used for DTN~\cite{Kohama_PRL_2011_DTN, Zhang_PRB_2013}:

\begin{equation}
\label{eqn:mass_renormalization}
\frac{m^*}{m} = \frac{H_{c2}}{4~H_{c1}} \left( 1 + \sqrt{1 + \frac{8~(H_{c1})^2}{(H_{c2})^2}} \right)
\end{equation}

where $m^*$ is the renormalised effective mass near $H_{\mathrm{c1}}$ and $m$ is the effective mass near the upper critical field. Using $\mu_0H_{\mathrm{c1}} = 0.17$~T and $\mu_0H_{\mathrm{c2}} = 8.20$~T gives $m^*/m = 24 \pm 3$. This value is much larger than the corresponding estimate for DTN, $m^*/m \sim 3.2$~\cite{Kohama_PRL_2011_DTN}, and reflects the extreme asymmetry of the Henmilite phase boundary.

Within this phenomenological interpretation, Henmilite therefore lies in an unusually fluctuation-dominated limit of field-induced triplon condensation: the lower critical field is exceptionally small, the parent state remains on the gapped side of the coupled-ladder instability, and the field-induced phase expands into a broad dome extending over several tesla. This makes Henmilite a particularly sensitive platform for testing how ladder quantum fluctuations, weak anisotropy, and proximity to zero-field order modify the conventional triplon-BEC picture.

\bibliography{bibliography}